\documentclass[fleqn,usenatbib,useAMS]{mnras}

\usepackage{graphicx}	
\usepackage{amsmath}	
\usepackage{amssymb}	
\usepackage{multicol}        
\usepackage{bm}		
\usepackage{pdflscape}	

\usepackage{url,times,graphicx,hyperref,amsmath,amsfonts,amssymb,aas_macros,color,epsfig,epstopdf}
\usepackage{booktabs}
\usepackage[utf8]{inputenc}
\usepackage{ae,aecompl}
\usepackage{subcaption}
\usepackage{xcolor}

\usepackage[T1]{fontenc}
\usepackage{ae,aecompl}

\usepackage{newtxtext,newtxmath}

\title[The unabridged MW satellite luminosity function]{The unabridged
  satellite luminosity function of Milky Way-like galaxies in
  $\Lambda$CDM: the contribution of  ``orphan'' satellites}

\author[Santos-Santos et al.]{
Isabel M.E.  Santos-Santos,$^{1}$\thanks{E-mail: isabel.santos@durham.ac.uk}
Carlos S.  Frenk,$^{1}$
Julio F. Navarro,$^{2}$
Shaun Cole,$^{1}$
John Helly$^{1}$
\\
$^{1}$Institute for Computational Cosmology, Department of Physics, Durham University, South Road, Durham, DH1 3LE, UK\\
$^{2}$Department of Physics and Astronomy, University of Victoria, BC V8P 5C2, Canada\\
}

\date{Accepted XXX. Received YYY; in original form ZZZ}

\pubyear{2024}

\begin{document}
\label{firstpage}
\pagerange{\pageref{firstpage}--\pageref{lastpage}}
\maketitle

\begin{abstract}
  We study the abundance, radial distribution, and orbits of luminous
  satellites in simulations of Milky Way-mass dark halos in the
  $\Lambda$CDM cosmology.  We follow the evolution of a halo from
    the Aquarius project and the formation of its ``maximal''
    satellite population with the GALFORM semi-analytic model of
    galaxy formation. This population consists of all subhalos able to
    form stars through efficient gas cooling before or after
    reionization, which effectively selects systems with peak circular
    velocities exceeding a critical threshold of roughly $\sim 15$-$20$ km/s.
    The total number of luminous satellites is sensitive to the
    assumed redshift of reionization, but the shape of the GALFORM
    satellite stellar mass function is robust, peaking at the stellar
    mass ($\sim 10^3\, M_\odot$) of a halo just above the critical
    threshold.  Subhalos are prone to artificial disruption
    in the tidal field of the main halo, with the number of surviving
    satellites increasing with resolution.  Even in the highest
    resolution simulation (Aq-L1, with particle mass
    $m_{\rm p}\sim10^3\, M_\odot$), a substantial number of satellite
    subhalos are disrupted, leaving behind ``orphan'' galaxies tracked
    in GALFORM by the subhalo's most-bound particle before disruption.
    When orphans are included (and the effects of tidal stripping on
    stars are neglected), all simulations that adequately resolve the
    critical threshold yield a converged maximal satellite stellar
    mass function. Most orphans were accreted early, are found in the
    central regions of the main halo, and make up roughly half of all
    satellites in Aq-L1.  Taking orphans into account there is
    no need to populate subhalos below the critical threshold with
    satellites to fit the radial distribution of Milky Way satellites,
    as has been argued in recent work.  Our model predicts that
    orphans dominate the ultra-faint population and that many more
    satellites with small apocentric radii should be detected in
    upcoming deep wide-field surveys.
\end{abstract}




\section{Introduction}\label{sec:intro}
The $\Lambda$-cold dark matter ($\Lambda$CDM) cosmological model
describes a Universe where structure grows through
gravitationally-driven hierarchical clustering of dark matter clumps
\citep[see][for a review]{FrenkWhite2012}.  Galaxies form as gas cools
and condenses at the centre of these dark matter ``halos''
\citep{WhiteRees1978,WhiteFrenk1991}. The resulting distribution of
dark matter structures, which resembles a cosmic web on large scales,
has proved to be consistent with the observed large-scale distribution
of galaxies at various epochs
\citep[e.g.]{Cole2005,Springel2006,Zehavi2011}.

$\Lambda$CDM predicts a steep halo mass function across the entire
mass range, from $\sim 10^{-6}\, M_\odot$ to $10^{15}\, M_\odot$
\citep[for a 100~GeV dark matter particle;]{Wang2012}.  Furthermore,
individual halos are filled with ``substructure'' consisting of the
surviving inner regions of halos that merged with the main halo
progenitor at earlier times. These subhalos also have a steep mass
function, with thousands of low-mass systems expected within the
virial radius of a host halo like that of our Milky Way (MW). In
contrast, only $\sim 66$ MW satellites have so far been identified, a
discrepancy that gave rise to the long-standing misconception of a
``missing satellites'' problem in $\Lambda$CDM
\citep[][]{Klypin1999,Moore1999}.

A simple solution to the ``missing satellites'' problem envisions
luminous subhalos (``satellites'', hereafter) forming only in halos
exceeding a ``threshold'' or ``critical'' mass, which means that
satellites are only able to populate the few most massive
subhalos. The astrophysical processes that prevent galaxies from
forming in halos below the threshold mass were understood years before
the ``missing satellites'' was hailed as a problem for CDM. 

Before the epoch of reionization, gas can condense and form stars only
in halos massive enough for atomic or molecular hydrogen to be able to
cool \citep{Tegmark1997,Benitez-Llambay2020}. After reionization, the
intergalactic medium is heated above $10^4$~K and gas is only able to
cool in halos above a circular velocity approximately corresponding to
the temperature of the ambient heated gas
\citep{Efstathiou1992,Thoul1996,Gnedin2000,Benson2002a}. These
mechanisms impose a characteristic, redshift-dependent halo mass scale
below which halos remain dark \citep{Benitez-Llambay2020}. 

Early work using semi-analytic models of galaxy formation applied to
merger trees constructed from the extended Press-Schechter formalism
showed how these mechanisms can be used to predict the abundance of
satellites in the Milky Way within the $\Lambda$CDM model  
\citep{Kauffmann1993,Bullock2000,Benson2002b,Somerville2002}. These
papers predicted that  MW-mass galaxies should have between 130 and
220  (90\% CL) satellites brighter than  $M_V=0$, although they were
written when only the 11 ``classical'' satellites of the MW were known
\citep{Benson2002b}. 

In the past couple of decades, the census of satellites in the MW halo
has expanded,  with discoveries facilitated first by SDSS and, later,
by deep imaging surveys such as DES,  Pan-STARRS1 and HSC
\citep[e.g.][]{Willman2005,Bechtol2015,Drlica-Wagner2015}. Currently,
66 satellites are known to orbit the MW but, accounting for the
selection function of these surveys
and the recent accretion of the LMC, 
\cite{Newton2018} estimated that
the MW should contain approximately  
$124^{+40}_{-27}$ (68 per cent CL) brighter than ${\rm M_V}=0$, in
broad agreement with the early predictions of \cite{Benson2002b}.   
 \citet{Nadler2020}  suggests an even larger estimate of $220\pm 50$.

In the past few years, cosmological simulations of increasing
sophistication and complexity have confirmed this
result. \cite{Sawala2016} showed that the APOSTLE cosmological
hydrodynamics simulations, based on the EAGLE code
\citep{Schaye2015,Crain2015}, reproduce the stellar mass function of
the classical satellites of the MW, for masses,
$M_{\rm *}>10^5\, M_\odot$, in agreement with the earlier
semi-analytic calculations. This result has also been echoed by other cosmological
hydrodynamical simulations
\citep[e.g.][]{Maccio2010,Simpson2018,Garrison-Kimmel2019}.  It is
therefore puzzling that, in spite of this extensive body of
theoretical work, references to the ``missing satellites'' problem in
$\Lambda$CDM are still found in the literature
\citep[eg.][]{Homma2019}.

Interestingly, the recent discovery of an increasing number of
ultra-faint satellite galaxies in the MW halo has led to the opposite
problem, namely that $\Lambda$CDM suffers instead from ``too many
satellites''. This new perceived challenge to $\Lambda$CDM is
predicated on the observation than there are more known satellites at
small distances from the centre of the MW than there are surviving
dark matter subhalos at those distances massive enough to form a
satellite in cosmological N-body simulations.

This apparent discrepancy led \cite{Graus2019} and \citet{Kelley2019}
to argue that subhalos of circular velocity as low as $\sim 7$ km/s
would need to harbour luminous satellites in order to reproduce the
observed radial distribution of MW satellites. This velocity is much
smaller than the  $\sim 12$ km/s corresponding to the critical
threshold mass discussed above \citep{Benitez-Llambay2020,
  Okamoto2009}. A similar problem has been highlighted for other
nearby galaxies by \citet{Carlsten2020}. 

However, it is important to point out that these studies did not fully take into account the extent to which the radial distribution of subhalos orbiting a MW-like halo is affected by the artificial disruption of subhalos that results from limited numerical resolution. Numerical convergence studies have shown that artificial disruption is important even for relatively massive subhalos in the highest resolution N-body simulations available \citep[see; e.g.,][]{vandenBosch2018,Errani2021}. A proper comparison of the radial distribution of satellites in the MW and other galaxies with $\Lambda$CDM predictions therefore requires careful consideration of these artificially disrupted subhalos.

In semi-analytic models of galaxy formation, satellites which have lost their dark matter subhalos are often called "orphans", and their existence has long been recognized as critical to explain the galaxy distribution in massive systems like galaxy clusters \citep[e.g.;][]{Guo2011}.

One of the main goals of this work is to assess the importance of
orphan galaxies and how they affect the predicted abundance, radial
distribution, and stellar mass function of satellite galaxies around
MW-mass halos. Our study makes use of the ``Aquarius A'' halo of the
Aquarius suite of N-body simulations \citep{Springel2008}. Aq-A is a
MW-mass halo that was simulated at 5 different resolution levels,
labelled L1 to L5; this allows a rigorous resolution study to be
carried out. The Aq-A-L1 simulation (particle mass of
$m_{\rm p}=1.7\times10^3$~M$_\odot$) remains the highest resolution
simulation of a MW halo ever carried out \citep[see also][for a
simulation of similar resolution]{Stadel2009}.
Our work is complementary to the statistical studies of this same halo by 
\citet{Han2016,He2024},
 who focus on disruption of dark subhalos at lower masses than  considered here.

To anticipate our main conclusions, we find that orphan galaxies are expected to be a major component of the satellite population, even in simulations with resolution as high as that of Aq-A-L1, where they make up roughly half of the total number of luminous satellites. Most orphan satellites are spatially concentrated, with many located tens of kiloparsecs from the host halo centre.
Our study highlights the importance of properly modelling
  orphan satellites to ensure reliable predictions, especially as
  upcoming surveys (e.g., DESI, LSST, Roman, Euclid) will probe the ultra-faint regime with unprecedented depth.

Our results here represent the maximal satellite population in a MW-sized halo, rather than a direct prediction for comparison with observations, especially in the MW’s inner regions. Indeed,
some of these orphan satellites may
suffer substantial 
stellar
mass loss due to tides, or disrupt entirely,
especially after the deepening of the potential well due to the
assembly of the baryonic component of the host galaxy. We do not
attempt to correct for these effects here, but intend to address them
in a future contribution. Our goal here is instead to provide a
prediction for the ``unabridged'' population of all satellites in
Aq-A; to check for convergence effects; and to assess which
$\Lambda$CDM predictions regarding the full satellite population,
including orphans, are most robust.

The paper is organized as follows. In Sec.~\ref{sec:methods} we
introduce the simulations, models and data used. In
Sec.~\ref{sec:orphans} we describe the treatment of orphan galaxies in
GALFORM.  In Sec.~\ref{sec:dwfgalform} we address dwarf galaxy
formation in the GALFORM model and introduce the satellite sample that
we will study. In Sec.~\ref{SecLumFun} we present our results for the
predicted luminosity function of satellites of MW-mass halos while in
Sec.~\ref{sec:radial} we present their radial distribution. In
Sec.~\ref{sec:orbits} we discuss the orbital evolution and present-day
orbital properties of satellites galaxies. Finally, in
Sec.~\ref{sec:conclu} we summarize our conclusions.

\section{Methods}\label{sec:methods}
\subsection{ The Aquarius-A dark matter halo}\label{sec:aq}

We use a simulated halo from the Aquarius project
\citep{Springel2008}, a suite of 6 dark matter-only MW-sized halos
simulated with the GADGET-3 code at 5 different resolution levels, L1
to L5.  The Aquarius project adopts a flat $\Lambda$CDM cosmological
model with parameters consistent with WMAP-5\footnote{The use of more
  recent estimates of the cosmological parameters is not expected to
  introduce any relevant quantitative changes to our results.}
\citep{Komatsu2009}: $\Omega_{\rm m}=0.25$; $\Omega_{\Lambda} = 0.75$;
$\Omega_{\rm bar} = 0.045$; $H_0 = 100\, h$ km s$^{-1}$ Mpc$^{-1}$;
$\sigma_8 = 0.73$; $h = 0.73$.

Specifically, we study the so-called Aquarius-A halo (hereafter Aq-A),
with virial\footnote{Virial masses are measured within a sphere with
  mean density 200 times the critical value for closure. Virial
  quantities are identified with a ``200'' subscript.} mass, 
$M_{200}=1.8\times 10^{12}$ M$_\odot$, virial radius, $r_{200}=245.7$
kpc, concentration, $c_{\rm NFW}=16.11$, and virial velocity, 
$V_{200}=179.4$ km/s, at $z=0$.

In this work we make use of resolution levels 1, 2, 4, and 5,
corresponding to particle masses
$m_{\rm p}/M_\odot=[1.7\times 10^3, 1.4\times10^4, 3.9\times 10^5,
3.1\times10^6]$ and Plummer-equivalent gravitational softening
lengths, $\epsilon/\textrm{pc}=[20.5, 65.8, 342.5,684.9]$,
respectively.  Aq-A-L1 remains to date the highest resolution
simulation of a MW-mass halo, with $\sim4.4$ billion particles in a
$100 \, h^{-1}$ Mpc cube and $\sim1.5$ billion of them within the
virial radius of the halo \citep[see also][]{Stadel2009}.  We refer the reader to
Table~1 of \citet{Springel2008} for more details on this particular
halo.

At $z=0$, the total halo mass and density profile of the Aq-A halo
show excellent numerical convergence at all radii across resolution
levels. A Navarro-Frenk-White profile \citep[][hereafter,
NFW]{Navarro1996,Navarro1997} provides an acceptable fit to the
density profile, as shown by \citet{Navarro2010}. Aq-A shows a steady
growth in virial mass without major late-time mergers, reaching
$10\%$ of its final total mass by cosmic time $t \sim 1$ Gyr,  and
$50\%$ by $t \sim 3.3$ Gyr.
For reference,  at $t=0.5(1)$ Gyr, the virial radius of the main progenitor was $\sim 5 (25)$ kpc (physical), 
the virial mass was $\sim10^{9.7} (10^{11})$ M$_\odot$,
 and  the virial velocity was $\sim 63 (160)$ km/s.

Subhalos in Aq-A were identified as self-bound substructures using the
SUBFIND algorithm \citep{Springel2001a} applied to the
friends-of-friends \citep[FOF,][]{Davis1985} group associated with the
main halo.  Embedded substructures survive within subhalos, as
expected from the hierarchical nature of structure formation in
$\Lambda$CDM.  A minimum of $20$ particles are required to define and
identify a subhalo at any time.

The substructure properties converge well between resolution levels
\citep{Springel2008}.  Specifically, at $z=0$, the differential mass
function of {\it surviving} subhalo converges well for masses
$<5\times10^8$ M$_\odot$ (above this mass there are very few massive
subhalos, resulting in more scatter), down to the $\sim 20$ particle
limit. This mass function can be adequately approximated by a
power-law with slope $-1.9$ \citep[see also][]{Gao2004}.  Similarly,
excellent convergence was also found in the cumulative subhalo maximum
circular velocity ($V_{\rm max}$) function, especially after
correcting the velocities for the gravitational softening length in
subhalos with scale radii comparable to the gravitational softening
\citep[see Eq.10 in][]{Springel2008}.

\citet{Springel2008}  also studied  the radial distribution of Aq-A-L1 subhalos as a function of mass.  They report that the number density profile of surviving subhalos follows a ``universal'' shape that is independent of subhalo mass and can be well described by an Einasto\footnote{The Einasto profile may be written as $\ln \rho/\rho_{-2}= (-2/\alpha)[(r/r_{-2})^\alpha - 1]$} profile with $\alpha=0.678$ and $r_{-2}=199$ kpc \citep[see also][]{Ludlow2009}. This profile is substantially shallower than the dark matter profile in the inner regions, and implies that the majority of surviving subhalos are found in the outskirts of the main halo. Even in our highest resolution simulation, no subhalos are found at $z=0$ within a distance of $r\sim10$ kpc from the halo centre.

\subsection{The GALFORM semi-analytic model}\label{SecGalform}

We populate the dark matter halos and subhalos\footnote{In practice,
  GALFORM takes the simulation's merger trees and particle data as
  input.} of Aq-A with luminous galaxies using the Durham
semi-analytical model for galaxy formation, GALFORM
\citep{WhiteFrenk1991,Kauffmann1993,Cole1994,Cole2000,Lacey2016}. 
GALFORM incorporates the many physical processes thought to govern the
physics of galaxy formation and evolution, namely, the shock-heating
and radiative cooling of gas in halos, star formation in galaxy disks
and in starbursts, feedback from supernovae and active galactic
nuclei, chemical enrichment of stars and gas,  and the
photoionization of the intergalactic medium, among others. 

The modelling of these processes relies on a set of parameters that
have been constrained to reproduce various observational
datasets. These include, at $z=0$, the galaxy luminosity functions in
different wavelengths, the HI mass function, and galaxy morphological
fractions, as well as the number and redshift distribution of massive
galaxies at high redshifts \citep[we refer the reader to][for further
details on the GALFORM model adopted here]{Lacey2016}.  GALFORM has
also proven to be successful at reproducing many observables for which
it was not tuned, such as the abundance of high redshift JWST galaxies
\citep{Cowley2018,Lu2024}, or the main properties of the ultrafaint
population of MW-like galaxies \citep[e.g.][]{Bose2018,Bose2020}.

The most relevant aspects of the GALFORM model in relation to this
study, which focuses on dwarf galaxies, are gas cooling, the impact of
the UV background responsible for the reionization of the Universe and
the impact of energy injected into the gas by supernovae. These
effects are included in GALFORM in an approximate way, with gas
cooling (and, hence, star formation) becoming ineffective in halos
whose virial masses are below the ``critical'' mass,
$M_{\rm crit}(z)$, for galaxy formation proposed by
\citet{Benitez-Llambay2020}. Before reionization, $M_{\rm crit}(z)$ is
computed from the virial temperature required to collisionally excite
neutral hydrogen. After reionization, the critical mass is set by the
minimum gravitational potential able to overcome the pressure of
photo-heated gas, bringing it to the halo centre and igniting star
formation. Calculation of $M_{\rm crit}$ before reionization was
already included in early versions of GALFORM \cite{Cole2000}; after
reionization, the ``$v_{\rm cut}$ '' prescription in those early
models has been replaced by an analytical fit to the
\citet{Benitez-Llambay2020} calculation.

We refer the reader to \citet{Benitez-Llambay2020} for details on the
critical mass calculation. For completeness, we list here the critical
mass at various redshifts: 
at $z=0$, $2$, $10$, $20$; $M_{\rm crit}=(5, 2, 0.08, 0.03) \times 10^9 \, M_\odot$,
assuming a redshift of reionization, $z_{\rm reion}=6$
\citep{Planck2020}. At that time the critical mass has a mild jump,
from
$\sim 10^{8.1}\, M_\odot$ to $10^{8.5}\, M_\odot$.
  We discuss this further in Sec.~\ref{sec:dwfgalform}.

Galaxy evolution in GALFORM varies if an object is a ``central'' or a
``satellite'' galaxy.  In central galaxies (i.e., those forming in the
main subhalo of a FoF halo), gas is allowed to cool and form
stars at any time. On the other hand, as soon as a galaxy infalls into
a more massive system and becomes a satellite (i.e., a subhalo of a
central galaxy), GALFORM assumes instantaneous ram pressure stripping,
meaning that all the diffuse gas of a subhalo halo is immediately
stripped away and added to the hot gas reservoir of the host halo.
Thereafter, gas is not allowed to accrete onto a satellite galaxy and
thus no further cooling occurs. Only the pre-cooled gas in the
satellite’s cold gas reservoir can continue to fuel star formation
after infall.
 
GALFORM (in the version adopted here) does not account for tidal
stripping of stars after a galaxy becomes a satellite. As a result,
the stellar mass of a satellite remains effectively fixed at the value
it had at infall (unless the galaxy retains some cold gas, which may
slightly increase the stellar mass as that gas forms new stars).

Finally, GALFORM models galaxy mergers by computing, at each timestep, a merging timescale for each galaxy with the central host galaxy.  Specifically, a modification of the analytical Chandraskhar dynamical friction formula is used \citep[see eq.6 in][]{Simha2017}, which emulates the results of cosmological simulations of galaxy formation.

\begin{figure*} \centering
  \includegraphics[width=0.49\linewidth]{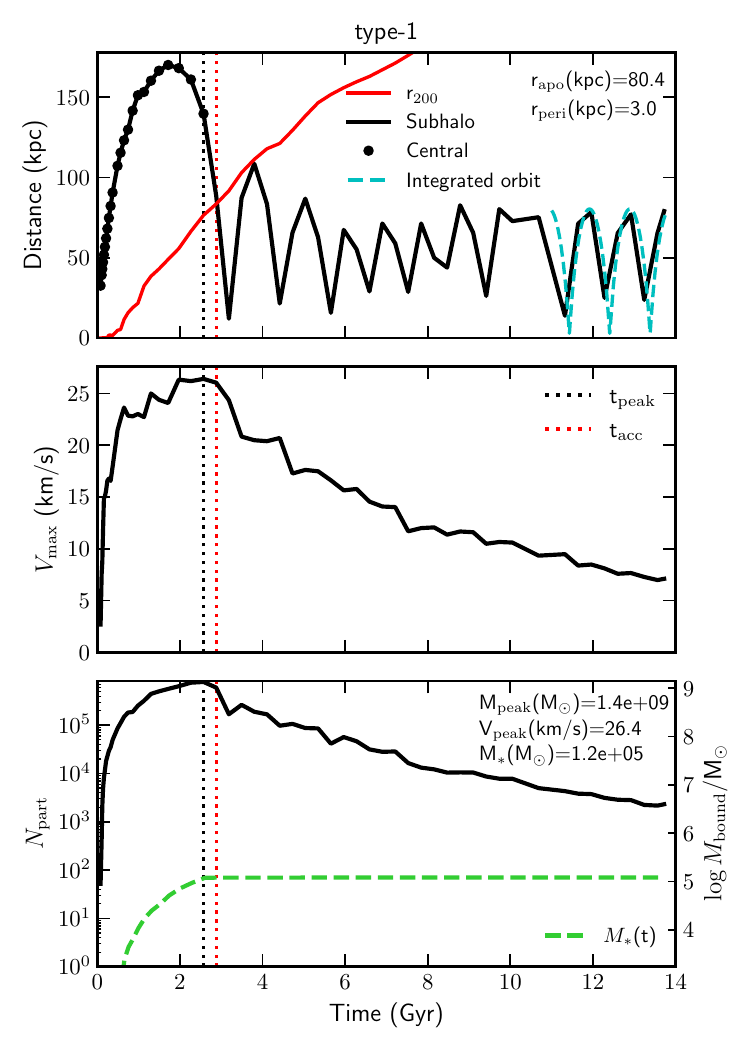}
  \includegraphics[width=0.49\linewidth]{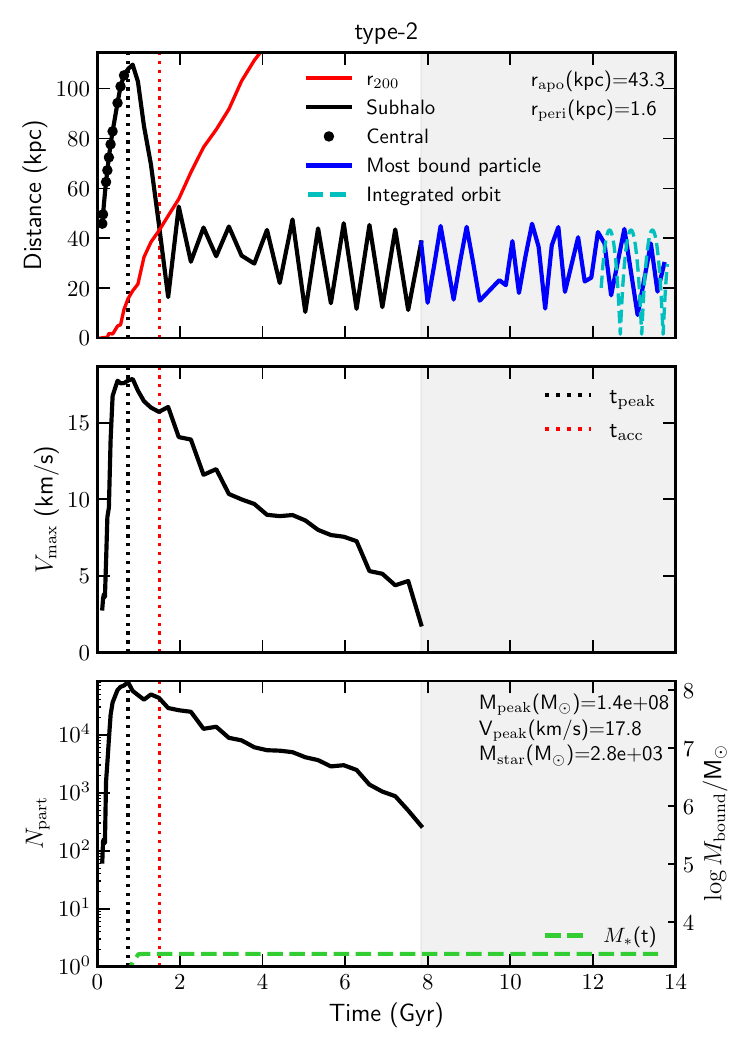}
  \caption{Evolution of example type-1 (left) and type-2 (right)
    satellites in Aq-A-L1.  The panels show the radial distance of the
    galaxy from the centre of the Aq-A halo (top), the maximum
    circular velocity $V_{\rm max}$ (middle) and the bound number of
    particles (and equivalent subhalo bound mass, $M_{\rm bound}$, see
    right y-axis in the bottom panels), as a function of the age of
    the Universe. The bottom panels also include, in green, the
    stellar mass of the subhalo, which remains essentially constant
    after accretion into the main halo as star formation
      ceases after accretion and stellar tidal mass loss is not
      considered in this work.  Solid black lines show satellite data
    from the {\sc SUBFIND} catalogue (while the subhalo is resolved).
    Black circles indicate when the galaxy is classified as the
    central of its {\sc FOF} group.  In the case of the type-2 galaxy,
    the background is shaded grey after the galaxy becomes an orphan
    and starts to be tracked via its most bound particle (shown with a
    solid blue line).  The red solid line shows the time evolution of
    the virial radius, $r_{200}$, of the Aq-A-L1 halo.  Vertical
    dotted lines indicate different characteristic times in the
    evolution of the satellite: the time when the circular velocity
    peaks ($t_{\rm peak}$, in grey) and the time of first infall into
    the main halo ($t_{\rm acc}$, in red); see text for their
    definitions.  The dashed cyan lines indicate the orbits integrated
    backwards in the gravitational potential of Aq-A-L1 from the $z=0$
    position and velocity.  }
 \label{FigEvoPlots}
\end{figure*}

\subsection{Orphan galaxies}\label{sec:orphans}

It is now generally recognized that in cosmological N-body simulations
many subhalos, especially those of low mass, fully disrupt in the
tidal field of the main halo. In early work, the tidal disruption of
these objects was often interpreted as indicating that subhalos
genuinely disrupt \citep[see e.g.][]{Hayashi2003}. Recent work,
however, has shown that numerical limitations play a substantial role,
artificially enhancing tidal disruption
\citep{vandenBosch2018}. Indeed, for $\Lambda$CDM halos, which
typically have a cuspy NFW-like density profile, the central cusp is
expected nearly always to survive the tides, even after intense and
prolonged tidal stripping \citep[][]{Errani2021}.  Galaxies within artificially
disrupted subhalos in N-body cosmological simulations are often
referred to in the literature as ``orphan'' galaxies.

GALFORM follows orphans after their subhalos fall below the
20-particle threshold and ``disrupt'' by tracking the most bound
particle of the subhalo at the last timestep when {\tt SUBFIND} is
able to identify it.  The position and velocity of this particle is
then assumed to be that of the orphan galaxy at later times. Orphans
may survive to $z=0$ or may be deemed to merge with the central galaxy
at a time determined by the Chandrasekhar dynamical friction timescale
computed just before the subhalo disrupts.

GALFORM thus classifies galaxies into three types: ``type-0'':
galaxies in resolved subhalos that are centrals in their FOF group;
``type-1'': galaxies in ``surviving", or resolved subhalos; and
``type-2'' or ``orphans'': satellite galaxies in disrupted or
unresolved subhalos. We shall make use of this nomenclature throughout
this paper.

\subsection{Orbital evolution of surviving and orphan satellites}

Fig.~\ref{FigEvoPlots} illustrates the orbital evolution of examples
of satellite galaxies of type-1 (surviving subhalo, left panels) and
type-2 (disrupted subhalo, right panels), chosen to highlight the main
characteristics of each type. The various panels show the time
evolution of the radial distance from the centre of the host halo
(top), of the subhalo maximum circular velocity, $V_{\rm max}$
(middle), and of the number of bound particles (or the equivalent
subhalo bound mass, $M_{\rm bound}$, see twin axis; bottom). The
temporal evolution of the virial radius of the main halo is depicted
in the upper panels with a solid red line.

Black lines indicate times when the subhalo is resolved as a surviving
self-bound system.  Times marked with a black circle indicate that the
galaxy is type-0 (i.e., central in its own FoF halo). In the case of
the type-2 galaxy, the background of the figure is shaded grey once
the subhalo becomes unresolved and the galaxy transitions into the
``orphan'' state.  From that moment onward, the satellite is tracked
by the most bound particle of its subhalo at the last timestep it was
resolved, and is shown by the solid blue line.

Vertical lines mark different characteristic times in the evolution of
a satellite.  The red dotted line indicates the time of infall into
the host halo, $t_{\rm acc}$, defined as the time when the galaxy's
radial distance first crosses the host halo's 
evolving
virial radius.
The black dotted vertical lines indicate the time, before infall, when
the number of particles associated with the subhalo reaches its
maximum, which we denote as $t_{\rm peak}$. The corresponding values
of $V_{\rm peak}$ and $M_{\rm peak}$ of these examples are listed in
the bottom panel legend.

In the two examples, $t_{\rm peak}$ occurs quite early ($\lesssim 3$
Gyr). For reference, in the implementation of GALFORM adopted here, the
time of reionization is only slightly earlier,
$t_{\rm reion}\sim 0.95$ Gyr ($z_{\rm reion}=6$). For the type-1
satellite, $V_{\rm peak}$ is $26$ km/s, while for the type-2 it is
$18$ km/s.  As for the bound mass, the type-1 and type-2 satellites
reach, at $t_{\rm peak}$, a maximum of $1.4\times10^9\, M_\odot$ and
$1.4\times10^8\, M_\odot$, respectively.

These satellites add a relatively small amount of mass to the main
halo, which, at $t=2$ Gyr, already had a virial mass of
$5\times10^{11}\, M_\odot$ and a virial radius of $\sim 50$ kpc. The
stellar mass assigned by GALFORM to each of these systems is shown by
the green curves in the bottom panel. These show that star
formation in GALFORM satellites typically ends shortly after infall
into the main halo; the stellar mass thus remains roughly constant
thereafter, without correction for possible tidal mass losses
\footnote{We use the tidal stripping framework of
    \citet{Errani2022} to estimate tidal mass losses in a forthcoming paper.}.

After infall, both satellites quickly settle onto fairly stable orbits
of constant apocentric and pericentric distances and relatively short
orbital times. This is because although the main halo keeps accreting
mass, little of that mass reaches the inner regions; indeed, the total
mass enclosed within $50$ kpc of the main halo remains essentially
constant after $t=3$ Gyr.

Note that the final apocentric distance is strongly correlated with the
virial radius of the main halo at the time of infall. For the type-1
example, the final apocentric distance is $\sim 80$ kpc, whereas the
virial radius at $t_{\rm acc}=3$ Gyr was $87$ kpc. The corresponding
numbers are $\sim 43$ kpc and $43.5$ kpc for the type-2 example, which
was accreted at $t_{\rm acc}=1.5$ Gyr. We shall see below that these
broad correlations are followed by all satellite systems identified
and tracked by GALFORM.

The evolution of both satellites shows clear signs of heavy tidal
stripping after accretion.  The bound mass of the type-1 satellite
decreases by more than two orders of magnitude after being accreted
into the main halo. The type-2 satellite loses at least as much, and
becomes unresolved after reaching a few hundred particles, when
SUBFIND loses track of it.

\subsection{Aq-A satellites}\label{sec:sample}

We focus on the satellite galaxies of the Aq-A host galaxy at $z=0$,
identified by GALFORM in four different realizations of the halo at
different resolutions. To recall, we define as satellites all subhalos
to which GALFORM has assigned a luminous component (i.e., stars) and
located within $300$ kpc from the centre of the main host galaxy
halo\footnote{This definition implies that not only galaxies
  classified as type-1 and 2 in GALFORM are considered satellites, but
  also any type-0 within $300$ kpc from the main halo centre.  We note
  that their contribution is, however, negligible: we find fewer than
  $2$ type-0s within $300$ kpc. These have been added to the type-1
  population.}. This limiting radius is slightly larger than the Aq-A
virial radius, $r_{200}=245$ kpc at $z=0$. We adopt $300$ kpc to
facilitate comparison with earlier work, which have conventionally
adopted $300$ kpc as a fiducial ``virial radius'' for the MW and M31
halos.

Using merger trees we are able to track all satellites back in time,
and measure their properties as a function of time. We shall focus on
$M_*$, the satellite stellar mass at $z=0$; its infall time,
$t_{\rm acc}$, defined as the first time it enters the
evolving
 virial radius
of the main progenitor; and on the orbital parameters at $z=0$, such
as pericentric and apocentric radii ($r_{\rm peri}$ and
$r_{\rm apo}$).
We note that a few satellites show recent infall times today, 
which could lead to future apocentres exceeding $r_{\rm 200}(t)$ (i.e. "backsplash" galaxies).  
Careful inspection of the orbital histories of all satellites however confirms 
that all luminous objects found within  $300$ kpc are indeed bound to the
 host halo, and no escaping galaxies are included in our satellite sample. 

We also track the satellite ``peak'' circular velocity ($V_{\rm peak}$),
defined as the maximum circular velocity, $V_{\rm max}$, at the time,
before infall, when the SUBFIND mass of a subhalo is maximal. We also
identify the time when a galaxy formed its first stars,
$t_{\rm fstr}$, according to GALFORM.  In GALFORM, gas can only begin
forming stars in galaxies that are centrals, so $t_{\rm fstr}$ in
general precedes $t_{\rm acc}$.

Special care has been taken to weed the GALFORM satellite list of
spurious objects, such as low-mass objects where stars have been
allowed to form only because the identity of the halo has been
temporarily swapped with that of a more massive neighbour, as well as
various other merger-tree artifacts in the low-mass regime. We use
this cleansed galaxy list to present, in the following section, our
results for the abundance, radial distribution and orbital
evolution of the full, unabridged luminous satellite population of the
Aq-A halo, with special emphasis on the contribution of orphan
satellites.

\begin{figure*} \centering \includegraphics[width=\linewidth]{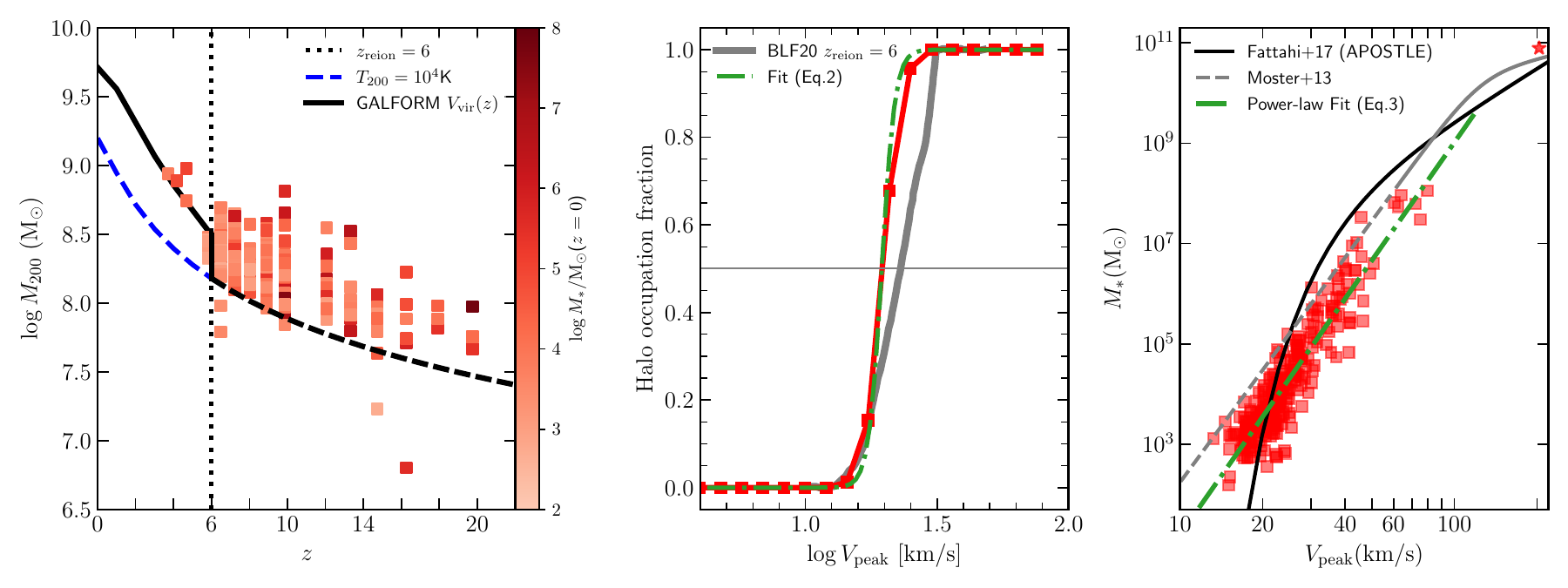}
  \caption{Galaxy formation in Aq-A-L1 according to GALFORM.
    \textit{Left:} redshift of formation of the first stars in a halo 
    ($t_{\rm fstr}$) as a function of halo mass,  $M_{200}$, at that time.  The
    redshift of reionization, $z_{\rm reion}=6$, is marked by a
    vertical line. Thick black lines illustrate the evolution of the
    critical mass, $M_{\rm crit}$ introduced by
    \citet{Benitez-Llambay2020}. Before reionization (dashed
    linetype), $M_{\rm crit}$ roughly tracks the ``hydrogen cooling
    limit'', i.e., the mass of a halo of virial temperature
    $T_{200}=10^4$ K.  After reionization (solid linestyle), $M_{\rm
      crit}$ equals the halo mass above which photoheated gas cannot
    remain in hydrostatic equilibrium and thus collapses. \textit{Middle:} halo
    occupation fraction (HOF) of halos in Aq-A-L1. 
    The HOF from \citet{Benitez-Llambay2020}, assuming $z_{\rm reion}=6$, is plotted in grey for comparison.  
    A green dot-dashed line shows our analytical fit to the HOF (Eq.\ref{eq:hof}).
    \textit{Right:} the GALFORM $M_*$-$V_{\rm peak}$ relation for Aq-A satellites.
    For comparison, the black curve shows the relation obtained in the
    APOSTLE simulations \citep{Fattahi2018} and the grey line shows
    the abundance-matching relation of \citet{Moster2013} extrapolated
    (in dashed linetype) to lower masses.  A star symbol shows the
    data point corresponding to the central host galaxy.  The green
    dot-dashed line shows the best power-law fit to the datapoints, with
    parameters given in Eq.~\ref{EqMstrVpFit}.  }
 \label{FigGalform}
\end{figure*}

\subsection{Milky Way satellites}\label{sec:MWdata}

We shall compare our results to the observational data of known Milky
Way satellite galaxies. This comparison is necessarily qualitative at
this stage, both because the census of MW satellites remains highly
incomplete, and because our results need to be modeified to include
stellar tidal stripping in both type-1 and type-2 satellites. As for
Aq-A, we shall consider as MW satellites all known dwarfs found within
$300$ kpc of the MW centre, which results in a sample of $66$
satellites;
see Appendix~\ref{sec:appMWdata} for details. While this limiting
radius corresponds to the virial radius of a halo of total mass larger
than current estimates for the MW
\citep[$M_{200}^{\rm MW}\simeq 10^{12}$ M$_\odot$, which corresponds
to $r_{200}=207$ kpc; see ][]{Cautun2020}, it is the radius commonly
used to define the satellite populations of the MW and M31 in the
literature. Had we only used satellites within $r_{200}=207$ kpc the
list would have not included Indus2, Bootes4, CanesVenatici1, Leo1,
Leo2, Cetus3, and Eridanus2, and would be reduced to $59$ MW
satellites.

We use positions (RA, dec) and distances $(m-M)$ data from the latest
update of the \citet{McConnachie2012} Nearby Dwarf Galaxy
Database\footnote{See
  \url{https://www.cadc-ccda.hia-iha.nrc-cnrc.gc.ca/en/community/nearby/}
  and references therein.}, and compute Galactocentric positions and
velocities assuming a Galactocentric distance for the Sun of
$R_\odot=8.29$ kpc, a circular velocity for the local standard of rest
(LSR) of $V_0=239$~km/s \citep{McMillan2011}, and a peculiar velocity
with respect to the LSR of $(U_\odot,V_\odot,W_\odot) = (11.1,
12.24,7.25)$ km/s \citep{Schonrich2010}. 

In Sec.~\ref{sec:orbits}, we shall compare our results to data for the
subset of MW satellites with available 3D kinematical information
($47$ objects).  We compute pericentric and apocentric distances using
the \citet{Cautun2020} MW potential, which assumes a virial mass for
the MW of $M_{200}=1.08^{+0.20}_{-0.14}\, M_\odot$ consistent
  with recent estimates \citep[e.g.][]{Callingham2019}.  Since this
potential is truncated at $300$ kpc we extrapolate the mass
distribution beyond this radius assuming an NFW mass profile with the
same $M_{200}$, when needed.  
The \citet{Cautun2020} potential
is static and includes different MW
 baryonic
  components (i.e. bulge, thin and thick stellar disc, HI disc,
  molecular gas disc, and diffuse gas halo), following parameters that
  best fit the latest Gaia rotation curve data \citep{Eilers2019}.  It
  also includes the effects of dark matter halo contraction from
  baryon effects.  We estimate orbital parameters by integrating the
  $z=0$ position and velocity data of MW satellites back in time for
  $\sim1$ Gyr. The integration neglects the effects of dynamical
  friction and time evolution in the host potential, which are likely
  negligible given the short integration time
  and the relatively small satellite halo masses.  
  The potential effects
  of a massive companion like the LMC are also not considered here
  \citep[see instead][]{Battaglia2022,Patel2020}.\footnote{We
      have compared our results using the \citet{Cautun2020} MW model
      with those of \citet{Li2021} and \citet{Battaglia2022} (who
      assume a lighter MW halo and the presence of the LMC in the case
      of the latter). While differences in the assumed MW
      halo mass and the inclusion or exclusion of the LMC naturally
      affect the precise values of the inferred MW satellite orbital
      parameters, all models broadly agree on the overall parameter
      space occupied by satellites on the $r_{\rm peri}-r_{\rm apo}$
      plane, which is what we shall focus on in this study.}  The derived
pericentric and apocentric distances are listed in
Table~\ref{tab:MWperiapo}.

\section{Results}
\subsection{Satellite galaxies in GALFORM}\label{sec:dwfgalform}

We begin our discussion by showing explicitly which halos are able to
harbour luminous galaxies in the implementation of GALFORM chosen in
this work. As discussed in Sec.~\ref{SecGalform}, before reionization,
stars can form only in halos massive enough for hydrogen cooling to
take place and, after reionization, in halos massive enough for
gravitationally collapsing cooling gas to overcome the pressure
induced by photo-heating. There is therefore a redshift-dependent
critical halo mass at all times, $M_{\rm crit}(z)$, above which stars
can form, as described in detail by \citet{Benitez-Llambay2020}. We
show this limiting mass as a function of redshift as the thick black
curves in the left-hand panel of Fig.~\ref{FigGalform}. Dashed
linetype corresponds to redshifts before reionization,
$z_{\rm reion}$; solid after $z_{\rm reion}$.

For comparison, a blue curve indicates the mass of halos with constant
virial temperature, $T_{\rm
  200}/$K$=(\mu m_p / 2 k_B)(V_{200}/$km/s$)^2=10^4$ K after
reionization.  The conversion from velocity to degrees K assume a mean
molecular weight of $\mu=0.597$ before and after reionization.

The square symbols show results for all Aq-A 
 $z=0$ satellites
at the time,
$t_{\rm fstr}$, of their first star formation episode (i.e., at the time
when they first appear in the GALFORM galaxy catalogue). We only
include in the plot
systems which were centrals (i.e., type-0s) at that time, 
to ensure  that their virial mass estimates,  $M_{200}(t_{\rm fstr})$,  are reliable.  (Indeed, 
if a subhalo has already become a satellite of the Aq-A host halo, 
its mass is ill-defined due to  tidal effects.)
Squares in Fig.~\ref{FigGalform} are coloured according to the stellar mass of
the system at $z=0$.
This figure shows that, as expected, luminous satellites mostly form
in halos which, at some time in their formation history, have had
virial masses exceeding the critical line. Note as well that the
great majority of satellites started forming stars before
reionization; only a handful form their first stars after $z_{\rm
  reion}=6$.

Most luminous systems start forming stars barely $\sim 0.3$ Gyr after
the Big Bang.  The last Aq-A-L1 satellite to begin forming stars does
so at $t_{\rm fstr}=7.5$ Gyr, i.e., $6$ Gyr ago. As noted by
\citet{Benitez-Llambay2020} and \citet{PereiraWilson2023}, this
explains why all MW satellites, regardless of mass, have at least some
fraction of their stars in a very old stellar population. In simple
terms, most halos able to form stars become eligible to do so very
early on. Note that, at the time of the formation of their first star,
most subhalos had virial mass exceeding $10^{7.5}\, M_\odot$. This is
not a result of limited numerical resolution but of the physics of
cooling gas. Indeed, such halos are well resolved in Aq-A-L1, with at
least $10^{4.2}$ particles.

The existence of a critical mass imposes a strong mass limit on the
Aq-A satellite population, which is best appreciated by considering
the ``occupation fraction'' as a function of $V_{\rm peak}$, i.e., the
fraction of Aq-A subhalos (identified at $z=0$) which harbour luminous
satellites, shown in the middle panel of Fig.~\ref{FigGalform}. Note
the sharp transition from completely ``dark'' subhalos to luminous
ones: all subhalos with $V_{\rm peak}<13$ km/s are dark, and all
subhalos with $V_{\rm peak}>28$ km/s have stars. These results are in
line with earlier work, particularly that by \citet{Okamoto2009} who
adopted a minimum halo circular velocity of $12$ km/s for hydrogen to
begin cooling and forming stars. They are also very close to the
analytic results of the \citet{Benitez-Llambay2020} model (depicted by
the grey line). For that model, the occupation fraction is $50\%$ at
$V_{\rm peak}=23$ km/s. In Aq-A-L1, the $50\%$ halo occupation is
reached at $V_{\rm peak}=19.5$ km/s.

GALFORM also predicts a strong dependence of the galaxy stellar mass
of a satellite on $V_{\rm peak}$, as shown in the right-hand panel of
Fig.~\ref{FigGalform}. Note the tight dependence of the galaxy stellar
mass on this measure of the depth of the subhalo potential well.
We note that $V_{\rm peak}$ may differ significantly from
$V_{\rm max}$ (the maximum subhalo circular velocity at $z=0$) as the
latter could have been strongly affected by tides. We also note that
$M_*$ in Fig.~\ref{FigGalform} refers to the stellar mass returned by
GALFORM, which does {\it not} include any loss of stars due to tidal
stripping. Tidal effects on the stellar component are likely to be
important for some satellites, especially those that evolve to become
an orphan at $z=0$. We intend to address this important issue in
detail in a future contribution.

The GALFORM satellite luminosity (or stellar mass) function will
depend strongly on the $M_*$-$V_{\rm peak}$ relation shown in the
right-hand panel of Fig.~\ref{FigGalform} so it is instructive to
compare it to the results of the cosmological hydrodynamical
simulations from the APOSTLE project \citep[solid
line,][]{Fattahi2018}, and of the abundance-matching model of
\citet[][]{Moster2013}. The differences between models highlights our
uncertain understanding of how dwarf galaxies populate low-mass
halos. Indeed, each of these models would yield significantly
different satellite luminosity functions for the same $V_{\rm peak}$
function, which is the true robust prediction of
$\Lambda$CDM. Although we will present results for the Aq-A satellite
luminosity function below, one should bear in mind this model
dependency before drawing strong conclusions.

In summary, the implementation of GALFORM adopted here allows luminous
satellites to form efficiently only in subhalos which exceed at some
point a threshold peak circular velocity of $\sim 15$ km/s.  The peak
circular velocity of a subhalo (typically reached just before first infall
into the main halo) determines, with little scatter, the satellite
stellar mass at the time of infall. Any modification to this model
would inevitably have a quantitative impact on our results regarding
the abundance of satellite galaxies, which we explore next.

\begin{figure}
\centering
\includegraphics[width=\linewidth]{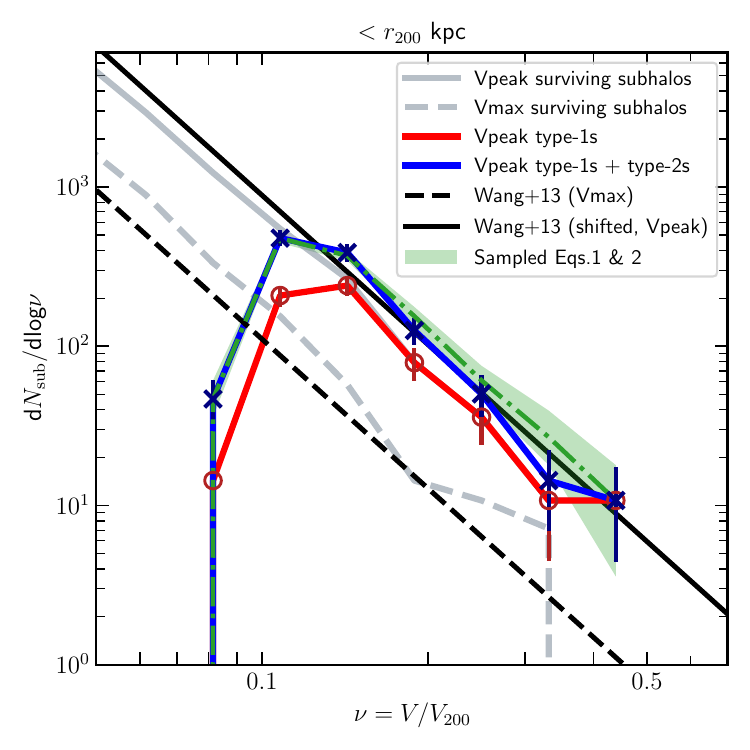}
\caption{ Differential number of galaxies (or subhalos) as a function
  of $\nu$, defined as the velocity in units of the main halo virial
  velocity (i.e., either $V_{\rm max}/V_{200}$ in dashed linetype or
  $V_{\rm peak}/V_{200}$ in solid linetype; see legend). Light grey
  lines show the differential subhalo $V_{\rm max}$ (dashed) and
  $V_{\rm peak}$ (solid) scaled functions for surviving subhalos in
  the dark matter-only Aq-A-L1 simulation. A black dashed line shows the
  scaled subhalo $V_{\rm max}$ function from \citet{Wang2012}.  
  Red and blue solid lines indicate the relations obtained for galaxy
  samples: type-1s only, and the total sample of satellites (type-1s +
  type-2s), respectively.  
  The solid black line is the
  \citet{Wang2012} relation, shifted by $0.29$ dex in $\log\nu$ to match
  the blue curve at large values of $\nu$ (see Eq.\ref{eq:wangvpk}).   
  The green shaded region shows the result of MonteCarlo sampling
    the subhalo $V_{\rm peak}$ function (Eq.\ref{eq:wangvpk})
  and filtering it by the HOF (Eq.\ref{eq:hof}). See text for details.
  }
 \label{FigVpeak}
\end{figure}

\subsection{The Aq-A subhalo peak-velocity function}\label{SecVpk}

As discussed in the previous subsection, the $V_{\rm peak}$ function
of {\it all} accreted subhalos 
in the Aq-A-L1 simulation
(which did not merge with the main
halo) underpins the GALFORM model predictions for the satellite
luminosity function. 
In this section, we study the $V_{\rm peak}$ function of Aq-A-L1 
subhalos and satellite galaxies, and introduce a simple model 
to estimate the GALFORM satellite luminosity function of  $\Lambda$CDM
host halos of arbitrary mass.

\citet{Wang2012} showed that the $z=0$ abundance of subhalos 
within $r_{200}$ of a host halo,  when plotted as a function of 
$\nu=V_{\rm max}/V_{200}$,  is independent of host halo mass (see
their fig.~3).  The cumulative subhalo $V_{\rm max}$ distribution may
be approximated by a simple power law: 
$\langle N_{\rm sub} \rangle (> \nu)=10.2 (\nu/0.15)^{-3.11}$ 
in the range $0.1<\nu<0.5$. Thus,
the total number of subhalos, expressed as a function of
$\nu_{\rm p}=V_{\rm peak}/V_{200}$, is also expected to be roughly
independent of host halo mass.

In order to make it relatively straightforward to scale our results for Aq-A to
other choices of the virial mass/radius/velocity of the main halo, 
we shall limit the discussion in this subsection
to subhalos that end up, at $z=0$, within the virial radius of the
main halo $r_{200}=245$ kpc.  Note that this choice of limiting radius is
different from the $300$ kpc 
used in the rest of the paper and
chosen to compare with the MW satellite population. 

We  begin the discussion by comparing, in Fig.~\ref{FigVpeak}, the
differential
$V_{\rm max}$ function for 
surviving subhalos in
Aq-A-L1 (dashed grey line) with the fitting formula
proposed by \citet{Wang2012} (thick dashed
black line), and find them in good agreement. Note that $V_{\rm max}$
refers to the maximum circular velocity of subhalos at $z=0$, and not
to $V_{\rm peak}$. 

The $V_{\rm peak}$ function of 
surviving
subhalos is shifted to higher
velocities relative to the $V_{\rm max}$ function, as shown by  the
thick  solid grey line in  Fig.~\ref{FigVpeak}. The velocity shift, of
order $\sim 60\%$, is nearly constant for all subhalos. The
largest subhalos in Aq-A have $\nu_{\rm p}\sim 0.4$-$0.5$, which
corresponds to $\sim 70$-$80$ km/s. This limit is only a factor of
$\sim 5$ larger than the minimum value of $V_{\rm peak}\sim 15$ km/s
needed for a subhalo to be able to harbour a luminous
galaxy. Therefore, the whole satellite luminosity function reflects
the abundance of subhalos over a very narrow range of $V_{\rm peak}$,
from $\sim 15$ to $\sim 80$ km/s. 

This is shown by the thick solid red line in  Fig.~\ref{FigVpeak},
which corresponds to the $V_{\rm peak}$ distribution of type-1
satellites. As expected, this tracks the solid grey line closely down
to $\nu_{\rm p}\sim 0.1$ (i.e., $V_{\rm peak}\sim 20$ km/s) where it
becomes exponentially truncated by the plummeting occupation fraction at
low $V_{\rm peak}$ (see middle panel of Fig.~\ref{FigGalform}). No
type-1 satellites are found with $\nu_{\rm p}<(15/179)=0.08$. 

Adding the ``orphan'' satellite population yields the thick blue
curve, which  leads to an increase in satellite numbers, especially at
$\nu_{\rm p}\sim 0.1$, near the truncation value. No orphans are found, either,
below  $\nu_{\rm p}=0.08$. 

We may approximate the total satellite $\nu$ function by sampling
  the $V_{\rm peak}$ relation (black  solid
  line) obtained by shifting the   \citet{Wang2012} function in
  velocity 
  to match the blue curve.
   In differential form, this function may be written as
\begin{equation}\label{eq:wangvpk}
    dN/d\log\nu = a (\nu/\nu_0)^b 
\end{equation}
with $a=660$, $b=-3.11$ and $\nu_0=0.11$. This function is
then filtered according to the HOF from this GALFORM model (see middle
panel in Fig.~\ref{FigGalform}). An analytical fit to the HOF (shown
with a green dashed line in that panel) is given by 
\begin{equation}\label{eq:hof}
 {\rm HOF}=   {(1  + \tanh( (x-x_0)/{x_1} ) ) } / {2}
\end{equation}
where $x=\log V_{\rm peak}/$km/s, $x_0=1.29$ and $x_1=0.05$.
The green line and shaded region in Fig.~\ref{FigVpeak} show the resultant mean and $\pm1\sigma$ uncertainty from 100 random realisations.   

\begin{figure}
\centering
\includegraphics[width=\linewidth]{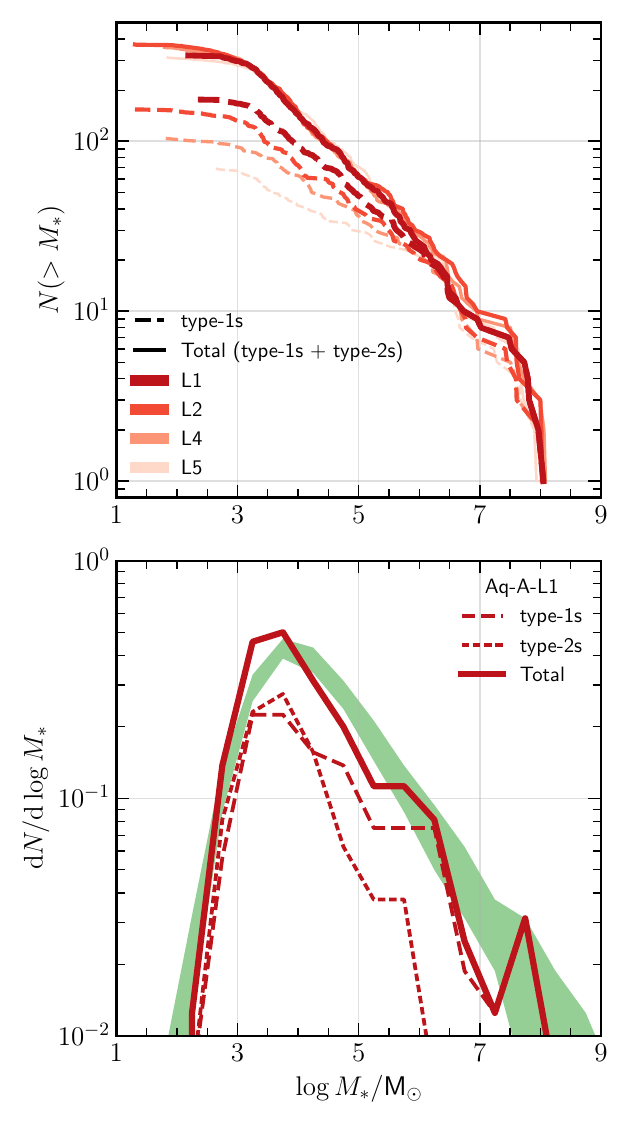}
\caption{ 
\textit{Top:} Satellite stellar mass function of Aq-A.  Lines of
different color/thickness correspond to GALFORM results for runs at
different resolution levels.  Dashed lines indicate the functions
obtained when counting only type-1 satellites (i.e.  satellites in
surviving subhalos). The solid lines show the total results, including
both type-1s and type-2s. 
\textit{Bottom:} Differential satellite luminosity function for
Aq-A-L1,  our highest resolution run.  We show separately the
contribution from type-1s, type-2s and the total. The green shaded area
shows the result of using the 
``filtered'' satellite subhalo $V_{\rm peak}$ function (see
Fig.\ref{FigVpeak}) together with the
 the $M_*$-$V_{\rm peak}$ relation to predict the satellite mass function. 
}
 \label{FigSatMstrF}
\end{figure}

These functions, together with a prescription relating  $V_{\rm peak}$
to $M_*$,  are enough to characterize the full, unabridged satellite
population of a MW-sized halo. The GALFORM $V_{\rm peak}$ -$M_*$
relation is well approximated by a power-law, 
\begin{equation}
M_*=9\times 10^{8}\, (V_{\rm peak}/100\, {\rm km/s})^{7.8}\, M_\odot , 
\label{EqMstrVpFit}
\end{equation}
shown by a green dashed line in the right-hand panel of
Fig. ~\ref{FigGalform}, with $\sim 0.4$ dex scatter in mass. 
 Using the above expressions,  we can analytically estimate the 
 differential abundance of satellites as a function of mass (see green shaded area 
 in the bottom panel of Fig.~\ref{FigSatMstrF}).

\begin{figure*}
\centering
\includegraphics[width=\linewidth]{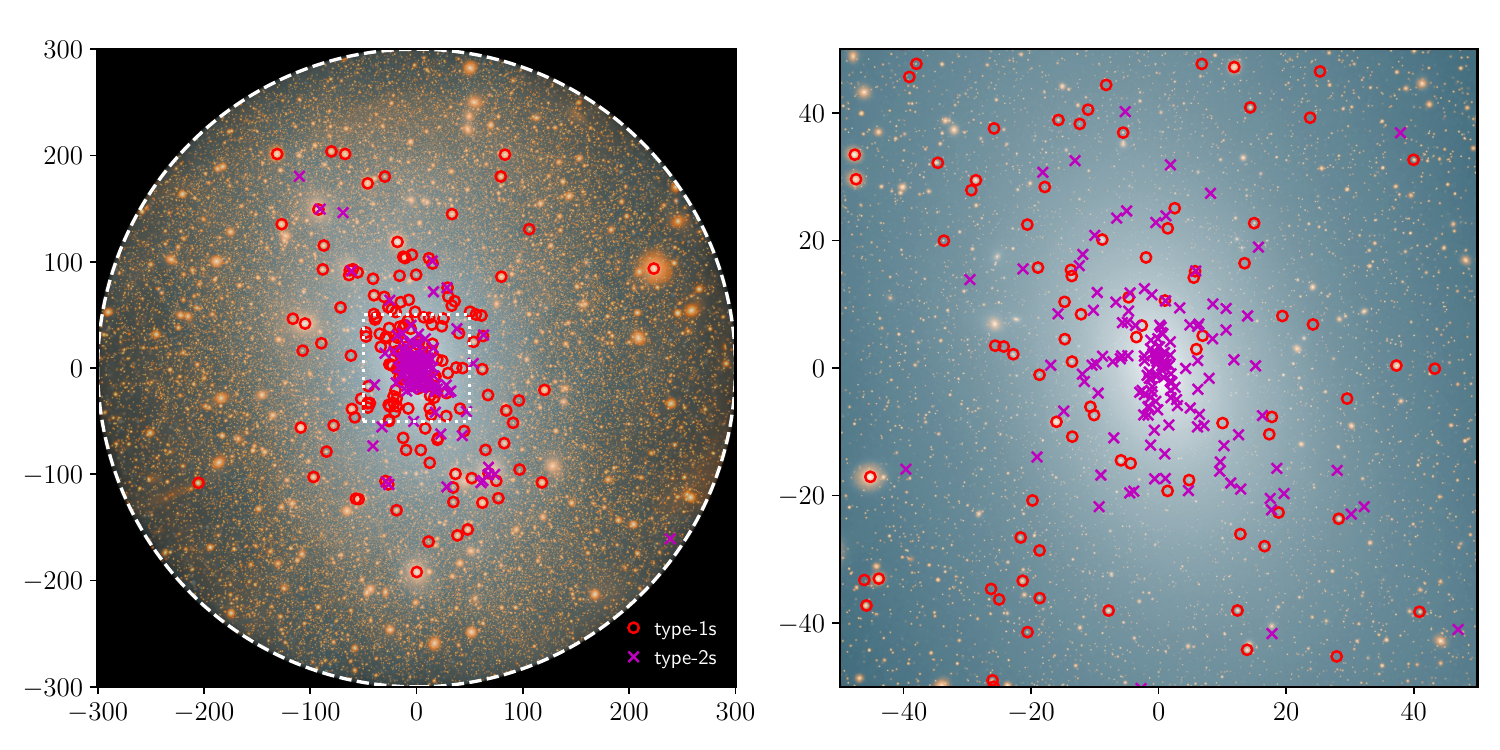}
\caption{ Illustration of the Aq-A-L1 halo and its satellites.  The
  background image shows a density plot where colour intensity
  correlates with dark matter particle velocity dispersion.
  Overplotted  are type-1 satellite galaxies (open circles) and type-2
  satellite galaxies (crosses).  A dashed circle marks our `virial'
  boundary ($300$ kpc) within which galaxies are defined as
  satellites. The right panel is a zoom-in of the inner $50$~kpc
  region around the centre of the Aq-A halo (marked in the left panel with a dotted rectangle). 
}
 \label{FigMap}
\end{figure*}

\subsection{The Aq-A satellite stellar mass function}\label{SecLumFun}

As discussed above, the satellite stellar mass function follows from
the peak velocity function explored in the previous subsection,
together with the tight $V_{\rm peak}$-$M_*$ relation shown in the
right-hand panel of Fig.~\ref{FigGalform}. We show the Aq-A satellite
stellar mass function within $300$ kpc at $z=0$ predicted by GALFORM
for the various resolution levels in Fig.~\ref{FigSatMstrF}. The top
panel shows the cumulative function, whereas the bottom panel shows
the same data in differential form, but only for the
highest-resolution run, Aq-A-L1.

The solid lines 
in the top panels of Fig.~\ref{FigSatMstrF} 
show results for \textit{all} satellites; i.e.,
surviving (type-1s) plus orphan (type-2s) systems. Different colours
indicate different resolution levels; the thickness of the curves
increases as the resolution improves. The contribution of type-1s
(surviving) satellites is shown by the dashed lines of matching
color. We recall that tidal stripping of stars is not modelled in the
implementation of GALFORM used for this study, so these mass functions
are best thought as corresponding to the ``infall'' or ``peak''
stellar mass function of Aq-A satellites.

Note the excellent convergence for the different resolution levels
when {\em all} satellites are considered (all solid lines are
basically on top of each other), although it is clear that the
contribution of orphan systems becomes more and more important as the
resolution becomes poorer. Type-1s are more prevalent at higher
resolution, with no sign of convergence, even in the highest
resolution runs. This is a clear indication that, even at the very
high resolution of L1, orphan satellites need to be considered to
account for the full satellite population of a MW-sized halo. When
orphans are included, all simulations converge. This is a reflection
of the relatively high halo mass ``threshold'' needed to host a
luminous system; indeed, halos of $V_{\rm peak}>15$ km/s
(corresponding to a virial mass of $\sim 10^{9}\, M_\odot$ at $z=0$)
are well resolved at all resolution levels of the Aq-A halo.

GALFORM predicts roughly $\sim 300$ Aq-A satellites with ``peak''
stellar masses above $10^3\, M_\odot$, and fewer than $\sim 400$
luminous satellites of all luminosities. For our highest resolution
run, {\em half of those satellites are orphans}, a fraction that
increases gradually at lower resolution.
We note that these numbers cannot be readily compared with extant data
for the MW satellite population, because they are particularly
sensitive to the virial mass of the halo, to the redshift dependence
of the mass threshold for galaxy formation, $M_{\rm crit}$, in the
model, as well as to possible corrections for tidal stripping, which
have not been considered here. Increasing or decreasing the halo
virial mass will affect almost linearly the predicted satellite
numbers. Varying the reionization redshift also has an effect, because
it leads to a change in $M_{\rm crit}(z)$, as we show in
Appendix~\ref{sec:appzrei}.

Although the total number of satellites is sensitive to modeling
details, one robust prediction is the presence of a well-defined peak
in the differential satellite stellar mass function, at the stellar
mass corresponding to that of a galaxy that forms in a subhalo with
$V_{\rm peak}\sim 20$ km/s, below which  the ``halo occupation
fraction'' drops dramatically (see middle panel in
Fig.~\ref{FigGalform}). In the GALFORM implementation we adopt here,
the peak is at  $\sim 10^{3.5}\, M_\odot$ (right-hand panel of
Fig.~\ref{FigGalform}), reflecting the fact that below $V_{\rm
  peak}=20$ km/s, more and more subhalos remain ``dark'', and those
that become satellites reach stellar masses $M_*>10^{3.5}\, M_\odot$. 

The peak at $10^{3.5}\, M_\odot$ does not seem to be a result of
resolution, since it is seen at all resolutions. Importantly, it is
seen in both type-1 and type-2 satellite mass functions. It may
therefore be considered a robust prediction of models that combine a
tight $V_{\rm peak}$-$M_*$ relation with a smoothly evolving
$M_{\rm crit}(z)$.  The actual location of the peak depends
sensitively on the $M_*$-$V_{\rm peak}$ relation.  However, for the
current GALFORM implementation, it does not change when varying
$z_{\rm reion}$, as we discuss in the Appendix~\ref{sec:appzrei}.
Identifying a well-defined peak in the stellar mass function of MW
satellites would help to put strong constraints on the validity of
this type of modeling.

Our conclusion is qualitatively similar to that of \citet{Bose2018},
who argued that cosmic reionization gives rise to a bimodal satellite
stellar mass function. A bimodal distribution can be seen in the
bottom panel of Fig.~\ref{FigSatMstrF}. The two peaks, at
$M_*\simeq 10^{3.5}\, M_\odot$ and $M_*\simeq 10^{6} \, M_\odot$, are
in the same location as in \citet{Bose2018} but the second peak is
less prominent in our study. The stronger bimodality in
\citet{Bose2018} reflects the sharp threshold, $V_{\rm peak}>30$ km/s,
for star formation after $z_{\rm reion}$ which alters the
$M_*$-$V_{\rm peak}$ relation.
The gentler evolution of $M_{\rm crit}(z)$ across reionization that
results from the \citet{Benitez-Llambay2020} model implemented here
reduces the prominence of the bimodality.

Finally,   in the bottom panel of Fig.~\ref{FigSatMstrF}, 
we present, as a green shaded region, the  satellite mass function 
predicted following the analytical model introduced in Sec.~\ref{SecVpk}.
This is obtained by combining the HOF-filtered  satellite subhalo 
$V_{\rm peak}$ function with the $M_*-V_{\rm peak}$ relation 
(see Eqs.\ref{eq:wangvpk}, \ref{eq:hof} and \ref{EqMstrVpFit}).
Specifically, the green area covers the $\pm1\sigma$ dispersion from
100 random realisations of this model.
As expected, the resulting analytical estimate closely reproduces
 the satellite distribution of the Aq-A-L1 simulation.

\subsection{The spatial distribution of Aq-A satellites}\label{sec:radial}

Fig.~\ref{FigMap} shows a randomly-oriented projection of the Aq-A-L1
halo and its satellite population at $z=0$. The background is an image
of the halo where the brightness of each pixel is given by the
projected dark matter density and the color intensity is determined by
the local velocity dispersion of dark matter particles. This rendition
emphasizes visually the presence of the smallest halos. 

The positions of satellite galaxies are overplotted:
surviving
(type-1) systems as open red circles, orphans (type-2s) as magenta crosses. Type-1
satellites lie on top of small subhalos, as expected, whereas type-2s
lack a corresponding dark matter counterpart.  
Aq-A-L1 resolves a large number of subhalos, much larger than the
total number of luminous satellites predicted by GALFORM. For
reference, in Aq-A-L1 there are $190,453$ resolved subhalos within
$300$ kpc at $z=0$, all with bound masses $>3\times10^4\, M_\odot$
(i.e., $> 20$ particles). About $\sim4000$ of these have, at present,
$V_{\rm max}>5$ km/s. As discussed in the previous subsections, only
$176$ of these surviving subhalos host luminous satellites (circles in
Fig.~\ref{FigMap}). A similar number of satellites ($144$) are orphans
(crosses).

The right-hand panel of Fig.~\ref{FigMap} shows a zoomed-in view of
the central $50$~kpc region. (The zoomed region is a slice that
excludes systems with $|z|>50$ kpc to minimize the inclusion of
distant satellites projected along the-line-of-sight.)  As expected,
most type-2s are located close to the centre of the halo, whereas
type-1s are found in general at larger radii.

As may be seen in Fig.~\ref{FigMap}, luminous satellites are more
centrally concentrated than the halo mass distribution. For example,
while the half-mass radius of the dark matter is $\sim 100$ kpc, the
radius that contains half of all luminous satellites is only $52$
kpc. As is clear from Fig.~\ref{FigMap} too, type-1s are much less
concentrated than type-2s. Half of all type-1s in Aq-A-L1 are found
within $82$ kpc, whereas the type-2 half radius is  just $18$~kpc. 
 
\subsection{Satellite number density profile}
\label{SecNumDensProf}

\begin{figure}
\centering
\includegraphics[width=\linewidth]{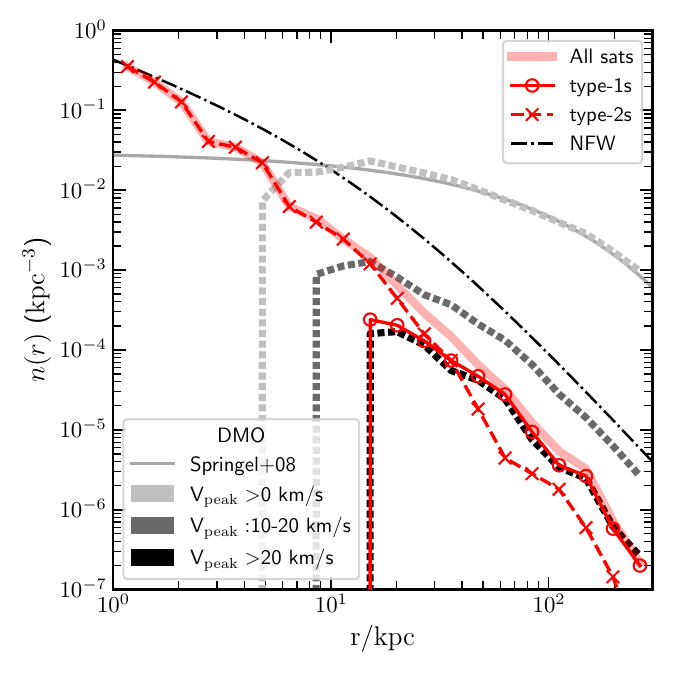}
\caption{ Number density profile of the satellite galaxies of
  Aq-A-L1. Red crosses correspond to orphan systems, red circles to
  satellites in surviving subhalos. The thick pink line is the sum of
  the two.  The dot-dashed curve, for comparison, shows an NFW profile
  with the same concentration as Aq-A-L1, arbitrarily scaled to match
  the same central value as the satellite profile.  The grey-scale
  dotted curves show the profiles of subhalos resolved in the
  simulation, in bins of $V_{\rm peak}$, as indicated in the legend.
  The grey line is the best-fit Einasto profile for all satellites from
  \citet{Springel2008}. }
\label{FigNumDensProf}
\end{figure}

The number density profile of all surviving subhalos in Aq-A-L1 has
been studied by \citet[][see their fig.11]{Springel2008}. In
particular, the shape of the number density profile was shown to be
nearly independent of subhalo mass and well fit by an Einasto profile
with parameter values $\alpha=0.678$ and $r_{-2}=199$~kpc \citep[see
also][]{Ludlow2009}.  This is a much shallower central dependence on
radius than that of the dark matter, which follows quite well an NFW
profile. We show this in Fig.~\ref{FigNumDensProf}, where the grey
dotted profile corresponds to all subhalos, regardless of mass,
identified at $z=0$ in Aq-A-L1. (The thin solid grey line is a
best-fit Einasto profile.) The contributions to this profile from
subhalos with $V_{\rm peak}>20$~km/s, and from subhalos with
$10<V_{\rm peak}/$km/s $<20$, are also shown with dotted grey lines, in
darker tones.

As expected, the number density profile of type-1 luminous satellites
(shown with red circles) is similar that of subhalos with
$V_{\rm peak}>20$ km/s. Orphan satellites (red crosses), on the other
hand, follow a much steeper profile, which crosses the type-1 profile
at $r\sim 30$~km/s. Inside this radius, type-2s dominate, whereas
outside type-1s dominate. Note, again that there are no type-1
satellites inside $10$ kpc; orphans, on the other hand, are found all
the way in, down to $\sim 1$~kpc from the main halo centre. We
remind the reader that the Aq-A-L1 simulation does not include the
gravitational potential of a central galaxy, nor tidal stripping of
stars, which are expected to affect the innermost regions of the
satellite number density profile.

In Aq-A-L1, type-1s dominate outside $30$ kpc, which implies that
surviving subhalos in N-body simulations at this resolution could, in
principle, be used to study the luminous satellite population outside
that radius. This limiting radius is, however, quite sensitive to
resolution: it grows to $69$, $102$ and $150$ kpc for resolution
levels 2, 4 and 5, respectively. So far, the highest-resolution
cosmological hydrodynamical simulations of MW-sized halos attempted
\citep{Sawala2016,Wetzel2023,Grand2024}
are comparable to level 2 (i.e., $m_p\approx 10^4\, M_\odot$), which
means that only satellites in the outer regions of the halo are
reliably resolved. A detailed treatment of orphans is therefore
critical in order to study the full luminous satellite population of
MW-like galaxies.

\begin{figure*}
\centering
\includegraphics[width=\linewidth]{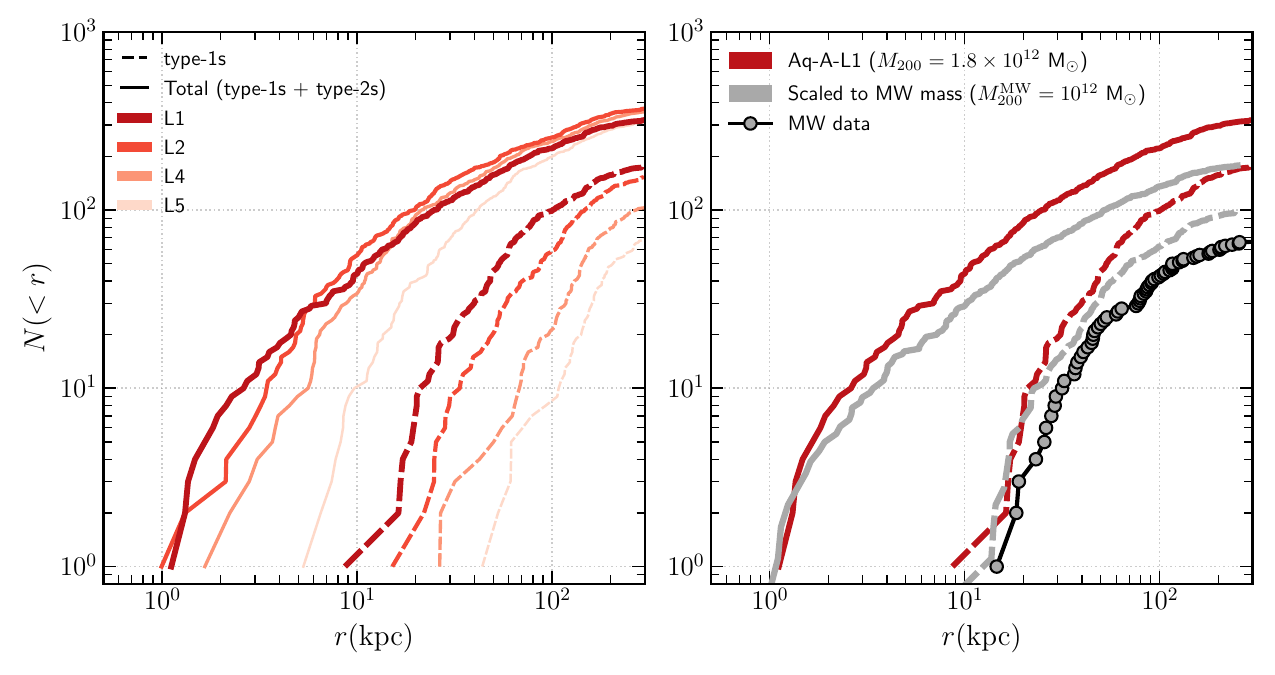}
\caption{ \textit{Left:} cumulative radial distribution of satellite
  galaxies in Aq-A at $z=0$. Different colours (and line thickness)
  correspond to different resolution levels.  Dashed lines show
  results counting only surviving subhalos (type-1s); solid lines show
  results for all satellites, i.e., type-1s plus type-2s.
  \textit{Right:} cumulative radial distribution of Aq-A-L1 satellites
  (dark red) compared to that of the currently known MW satellites
  (grey circles).  The latter list is likely to be highly incomplete.
  Since the number of satellites scales roughly with the mass of the
  host, Aq-A-L1 results from the left panel have been scaled by the
  estimated total virial mass and radius of the MW, for a more
  appropriate comparison (grey curves).  We use
  $M_{200}^{\rm MW}=10^{12}$ M$_\odot$ \citep{Cautun2020} and the 
  corresponding $r_{200} =207$~kpc that results from assuming an
  NFW density profile of average concentration.  }
 \label{FigRadDist}
\end{figure*}

\subsection{Cumulative radial distribution}

More detail on the satellite spatial distribution is provided in
Fig.~\ref{FigRadDist}, where we plot their cumulative radial
distribution at $z=0$. The left-hand panel shows results for
simulations of different resolution; the right-hand panel highlights
results for the highest resolution simulation and compares them to
known MW satellites.

As in Fig.~\ref{FigSatMstrF}, thick solid lines correspond to the full
satellite distribution (i.e., type-1s plus type-2s); dashed lines show
the radial profile of type-1 satellites only. Few type-1s are present
in the innermost regions of the main halo; even for L1 there are no
type-1 satellites inside $10$ kpc from the centre, a radius that
increases as the resolution decreases. In L5, there are no type-1s
within $50$ kpc from the halo centre.

Type-2s, on the other hand, are heavily biased towards the halo
centre, meaning that there is no shortage of satellites expected in
the inner regions of a MW-like halo once orphans are accounted for.
Within a radius of $10$ kpc, we find approximately $\sim 50$ Aq-A-L1
satellites, all of which are type-2s.

Overall, the radial distribution of satellites converges less well
than the satellite mass function, suggesting that it is much harder to
track subhalo orbits than it is to assign them a stellar
mass. Reassuringly, though, results for the two highest resolution
simulations (L1 and L2) are practically indistinguishable from each other,
suggesting that the radial distribution of all luminous satellites has
converged.

Note that the Aq-A-L1 satellite population is quite different from
that of known MW satellites, both in total number and in radial
distribution. As seen in the right-hand panel of
Fig.~\ref{FigRadDist}, whereas GALFORM predicts a total number of
$\sim 300$ satellites for Aq-A, only $\sim 66$ are known for the Milky
Way (grey circles in Fig.~\ref{FigRadDist}). Furthermore, GALFORM
predicts that half of all satellites are to be found within $\sim 50$
kpc, whereas half of known MW satellites lie outside $80$ kpc.  No MW
satellites have so far been found within $20$ kpc from the MW centre,
whereas GALFORM predicts roughly $85$.  Correcting for the roughly
$\sim 1.8\times$ higher halo mass of Aq-A than of the MW results in the
grey curves in the right-hand  panel of Fig.~\ref{FigRadDist}, but does not
change qualitatively the noted discrepancy.

These differences are not unexpected, since the known population of MW
satellites is almost certainly rather incomplete, and their radial
distribution is at present poorly constrained \cite[][and references
therein]{Newton2018}. On the other hand, the GALFORM Aq-A-L1
prediction likely overestimates the total number of satellites as well
as the central concentration of their radial distribution. Indeed,
tidal stripping (enhanced by the presence of the central galaxy) would
likely remove a fair fraction of orphans from the population,
especially those orbiting in the innermost regions of the main halo.

Although it is difficult to draw strong conclusions without explicitly
correcting for these effects, there appears to be, in principle, no
insurmountable difficulty in reconciling the known MW satellite
population with the Aq-A-L1 results. In particular, there is no
shortage of luminous satellites expected in the inner regions of the
MW halo; even without considering orphans, the radial distribution of
Aq-A-L1 satellites already provides a reasonable match to the MW known
satellite radial distribution (see right-hand panel of
Fig.~\ref{FigRadDist}).

This conclusion is apparently at odds with those of \citet{Kelley2019}
and \citet{Graus2019} who argued that too few luminous satellites
would remain near the halo centre once the enhanced tidal effects
resulting from the growth of the central galaxy are taken into
account. This view led these authors to argue that luminous satellites
should form in halos much less massive than the critical mass
threshold we have applied here.

Their conclusions, however, were based on the analysis of only type-1
satellites in simulations with lower resolution than ours (their
particle mass is $3\times 10^4\, M_\odot$ compared with
$1.7\times 10^3\, M_\odot$ for Aq-A-L1). As Fig.~\ref{FigRadDist}
makes clear, the presence of orphan satellites (which dominate the
innermost satellite population) challenges such conclusions. A proper
treatment of the tidal evolution and survival of satellites in the
innermost regions, as well as improved constraints on the full
inventory and radial distribution of the MW satellite population, are
needed to make further progress on this problem.

\begin{figure*}
\centering
\includegraphics[width=0.40\linewidth]{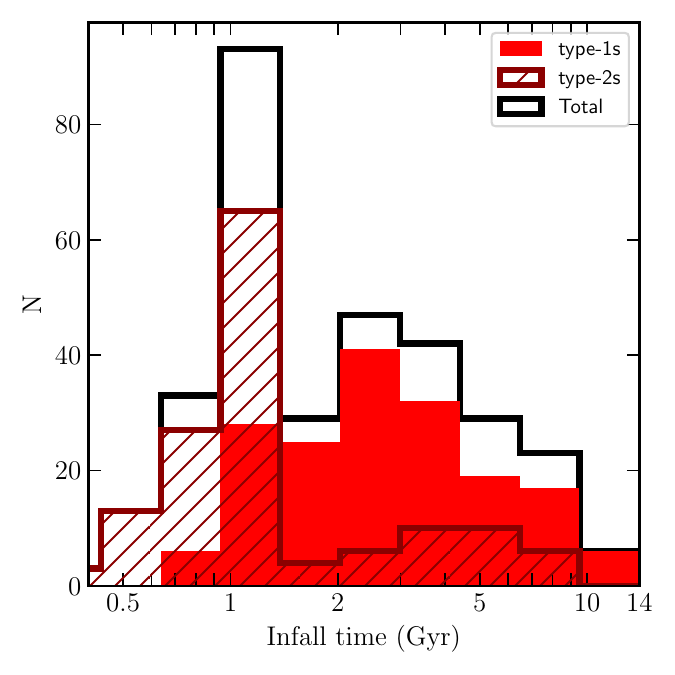}
\includegraphics[width=0.40\linewidth]{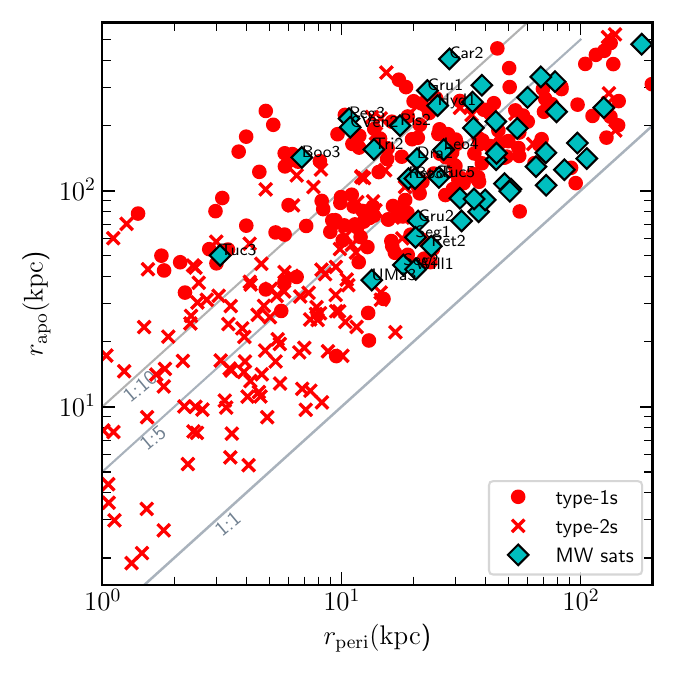}\\
\includegraphics[width=0.40\linewidth]{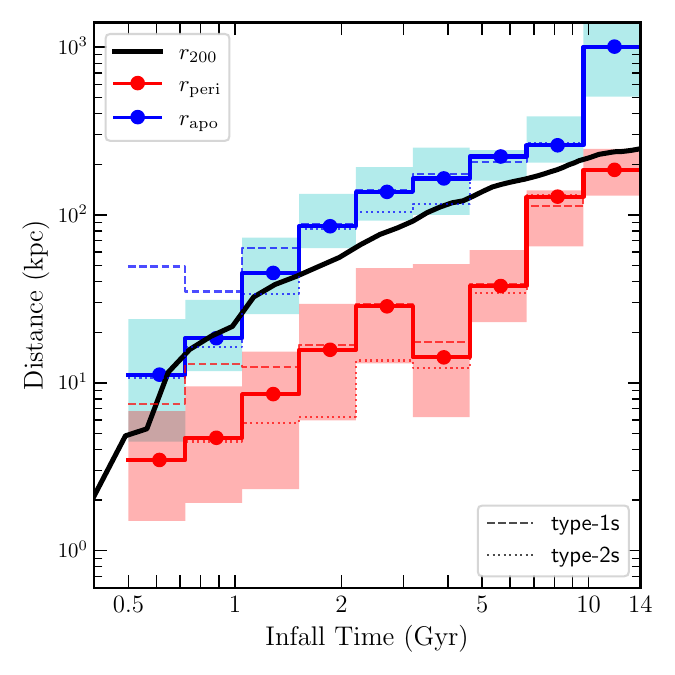}
\includegraphics[width=0.40\linewidth]{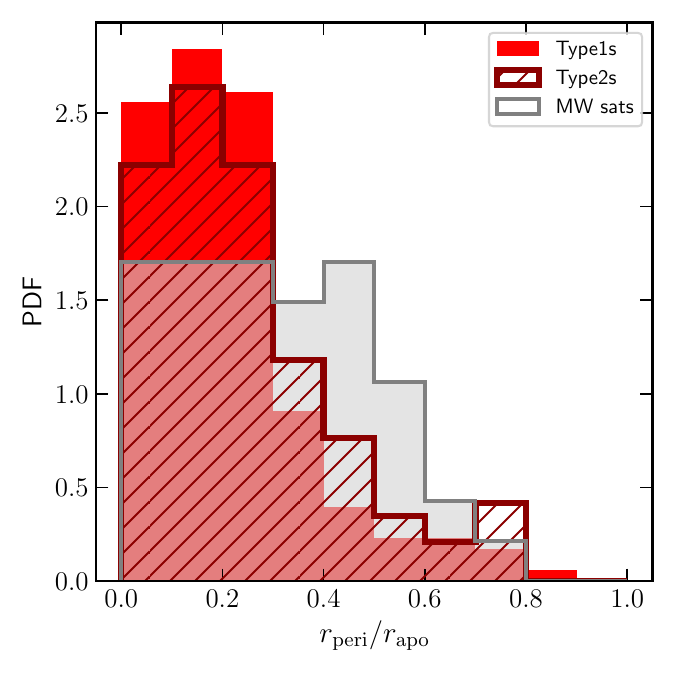}
\caption{ Orbital characteristics of Aq-A-L1 satellites.
  \textit{Top-left:} distribution of infall times for all satellites,
  including both type-1s and type-2s.  \textit{Top-right:} pericentric
  {\em vs} apocentric distances at $z=0$ for type-1s, type-2s and MW
  satellites.  Thin grey diagonal lines indicate different
  eccentricities, defined as $r_{\rm peri}$:$r_{\rm apo}$.
  \textit{Bottom-left:} median and $\pm1\sigma$ scatter in pericentric
  (red line) and apocentric (blue line) distance of satellites in bins
  of infall time.  A black line shows the evolution of the virial
  radius with time.  \textit{Bottom-right:} normalized distribution of
  eccentricities for type-1s, type-2s and MW satellites.  }
 \label{FigOrbits}
\end{figure*}

\subsection{Orbital properties of Aq-A satellites}\label{sec:orbits}

We have already discussed in Fig.~\ref{FigEvoPlots} the orbital
properties of two example satellites in Aq-A-L1, a type-1 and a
type-2. We now discuss the orbital properties at $z=0$ of all luminous
satellites identified in the simulation.

The distribution of infall times for the full satellite sample is
shown in the upper-left panel of Fig.~\ref{FigOrbits}.
Type-2s are typically accreted earlier than satellites that
survive as bound, resolved subhalos until $z=0$. This is because, as
in the examples shown in Fig.~\ref{FigEvoPlots}, satellites accreted
earlier have, at the present time, smaller apocentric and pericentric
radii, as well as shorter orbital times. Unfortunately, computing
accurate orbital parameters from the sparse snapshot data stored for
Aq-A-L1 is challenging, so we resort to using the $z=0$ position and
velocity of each satellite to integrate backwards in time, keeping the
gravitational potential of the main halo fixed. 
Specifically,  the gravitational potential is modelled from the 
mass distribution of the Aq-A-L1 halo at $z=0$ assuming spherical symmetry,
and we integrate for only $1.5$ Gyr back in time.
For the  mass profile we consider all particles associated with the $z=0$ Aq-A  halo
as defined by  Subfind,  reaching out to $640$ kpc.  
Two example orbital
integrations are shown by the cyan dashed lines in
Fig.~\ref{FigEvoPlots} (top panels). 
Pericentre and apocentre distances for each satellite are 
determined by identifying the radii at which the 3D radial velocity 
of the satellite changes sign.

These integrations are seen to reproduce the satellite orbits quite
well, and may therefore be used to estimate present-day $r_{\rm peri}$
and $r_{\rm apo}$ for all satellites. We show these as a function of
infall time in the bottom-left panel of  Fig.~\ref{FigOrbits}. As
anticipated when discussing Fig.~\ref{FigEvoPlots}, the apocentric
distance of a satellite is, on average, a good measure of the virial
radius of the main halo at the time of the satellite's first
infall. This explains the offset between the typical infall times of
type-1s and type-2s: satellites accreted earlier (generally, orphans)
have smaller apocentric and pericentric radii, are subject to much
stronger tidal stresses and complete more orbits than systems accreted
later, which have a greater chance to survive bound until $z=0$. 

The distribution of infall times is double-peaked, a result that may
be traced to a fairly major accretion event which nearly doubled the
main halo mass at $t\sim 3$ Gyr. The accreted halo brought its own
satellites, which settled onto orbits with higher eccentricity than
the average. This may be seen by the dip in pericentric radii in the
bottom-left panel of Fig.~\ref{FigOrbits}, which coincides with the
second peak in the infall time distribution.

The bottom-left panel of Fig.~\ref{FigOrbits} also indicates that
there should be a broad correlation between pericentric and apocentric
radii at $z=0$, which we show in more detail in the top-right panel of
the figure.  Here, solid circles represent type-1s, while crosses
depict type-2s.  Diagonal grey lines indicate different eccentricities
(i.e., different values of the ratio $r_{\rm peri}/r_{\rm apo}$ - 1:1,
1:5, 1:10). The typical eccentricity of Aq-A satellites is roughly
1:5, although the distribution is broad, and contains a number of
satellites in highly eccentric, plunging orbits with fairly small
pericentres.

As expected from this discussion, the orbits of type-1s and type-2s
differ mainly in their apocentric radii. Indeed, few orphans have
$r_{\rm apo}>30$ kpc, and, conversely, there are few, if any,
surviving satellites with $r_{\rm apo}<30$ kpc. This boundary is the
radius that separates the region where type-1s dominate over type-2s,
as discussed in Sec.~\ref{SecNumDensProf}.

It is interesting to compare these results with current estimates of
the orbital parameters of MW satellites for which full kinematic
information is available (see Sec.~\ref{sec:MWdata} and Sec.~\ref{sec:appMWdata}). 
The MW data are shown in the top-right panel of Fig.~\ref{FigOrbits} as cyan diamonds,
accompanied by the satellite's name if $r_{\rm peri}<30$ kpc.

Although most MW satellites overlap in this parameter space with
Aq-A-L1 type-1 satellites, there are some notable differences
too. Compared to Aq-A-L1, the MW has a dearth of satellites with small
pericentric radii, $r_{\rm peri} < 10$ kpc. Such objects are not rare
in Aq-A-L1, but there is only one MW satellite with such small
pericentre, Tucana 3
\citep{Drlica-Wagner2015,Shipp2018,Errani2023}. Although there is
still some debate as to whether Tucana 3 is actually a dwarf galaxy or a
star cluster, it exhibits prominent tidal tails and is in the process
of being tidally disrupted. This is a hint that many satellites with
$r_{\rm peri}<10$ kpc may have been disrupted by tides, a process that
is not included in the current study.
Likewise, the Aq-A halo potential used in this work lacks the presence 
of a central baryonic disk, whose additional mass is expected to further 
influence the simulated satellites' orbits.

There is also the possibility that MW satellites with small
$r_{\rm peri}$ could be missing from the current inventory. If these
galaxies are very faint (perhaps because they have been heavily
stripped), they would be very difficult to identify. One example is
the recently detected dwarf galaxy Ursa Major 3, a remarkable system
with an estimated stellar mass of only $\sim 14\, M_\odot$
\citep{Smith2024}. Although the nature of this object is still
debated, its vulnerability to tidal stripping favours its
interpretation as a dwarf galaxy \citep{Errani2024}. Objects like this
are easy to miss, and call into question how complete our current
inventory of faint MW satellites really is.

The bottom-right panel in Fig.~\ref{FigOrbits} compares the
eccentricity distribution of MW satellites with that of
Aq-A-L1. Athough generally consistent, there is some evidence that MW
satellites have slightly less eccentric orbits than those in Aq-A-L1. 
This difference may  be influenced by the individual merger histories 
of the MW and the Aq-A halo, as satellites accreted during past mergers 
could have settled onto orbits with different eccentricities.  Nonetheless, we
note that these distributions may be greatly modified once tidal
stripping processes are included, so we refrain from drawing strong
conclusions from this comparison.

\section{Summary and Conclusions}
\label{sec:conclu}

We have used the Aquarius suite of dark matter-only zoom-in
cosmological simulations to study the abundance, radial distribution,
and orbital properties of satellite galaxies in MW-sized halos formed
in a $\Lambda$CDM universe. We focus on the Aq-A halo, which was run
at various numerical resolutions, enabling an assessment of the
sensitivity of our results to numerical limitations. The simulations
were coupled with the GALFORM semi-analytical model of galaxy
formation to track subhalos able to form stars which accrete into the
main halo at various times. This procedure allows us to track every
single luminous satellite expected to form and fall into this halo.

The GALFORM implementation we use assumes that, as discussed by
\citet{Benitez-Llambay2020}, galaxy formation can only take place in
dark matter halos whose mass exceeds a redshift-dependent critical
value, which, before reionization, is equal to the mass at which
atomic hydrogen cooling becomes effective and, after reionization, is
equal to the mass above which gas cannot remain in hydrostatic
equilibrium in the presence of a photo-ionizing background.  In
practice, this criterion implies that galaxy formation is restricted
to halos with peak circular velocity exceeding $ 13$ km/s. In
terms of $V_{\rm peak}$, the transition between luminous galaxies and
``dark'' subhalos is extremely sharp: $95\%$ of halos with
$V_{\rm peak}=25$ km/s are luminous, but only $5\%$ of $15$ km/s halos
harbour luminous galaxies. Below $V_{\rm peak}=13$ km/s all halos
remain ``dark'', whereas $100\%$ of halos above $28$ km/s host
luminous components.

GALFORM also tracks all luminous subhalos (i.e., ``satellites'')
regardless of whether their dark matter component survives the tidal
field of the main halo as a self-bound entity. ``Orphan'' satellites
whose subhalos have artificially disrupted are an important component,
contributing about half of the full, unabridged satellite population
predicted by GALFORM. Orphan satellites have typically earlier infall
times, smaller pericentric and apocentric radii, and shorter orbital
times than those in surviving subhalos.

Although the inclusion of orphans is a critical improvement of our
study over some earlier work, our current implementation neglects the
effects of tidal stripping on the stellar component of satellites, as
well as the enhanced tidal field due to the assembly of the central
galaxy. These two effects are expected to reduce significantly the
number of satellites on orbits with small pericentric radii and short
orbital times. Therefore, our results are best thought of as the
``maximal'' satellite population expected in a MW-sized halo, and not
an actual prediction to be directly compared with observational data,
particularly in the inner regions of the MW. 

Our main findings may be summarized as follows.
\begin{itemize}

\item A total number of $\sim 300$ satellites within $300$~kpc is
  predicted by our model for Aq-A, a halo with virial mass,
  $M_{200}=1.8\times 10^{12}\, M_\odot$, which is likely about twice
  as massive as that of the Milky Way \citep{Cautun2020}. This is
  therefore a firm upper limit to the {\it total} number of MW
  satellites predicted by our model. This number depends linearly on
  the assumed MW virial mass, and is sensitive to the assumed
  reionization redshift and to the details of the evolution of the
  critical mass, $M_{\rm crit}(z)$. About half these satellites are
  ``orphans'' in our highest resolution simulation, Aq-A-L1, which
  contains roughly $1$ billion dark matter particles within the virial
  radius of the main halo.

\item GALFORM predicts a tight relation between $V_{\rm peak}$ and
  $M_*$. Because of the sharp transition in $V_{\rm peak}$ between
  ``luminous'' and ``dark'' subhalos, the satellite stellar mass
  function is expected to peak at the stellar mass that corresponds to
  $V_{\rm peak}\approx 15$~ km/s, i.e., $M_*\sim 10^{3.5}\,
  M_\odot$. This implies a sharp downturn in the number of satellites
  with stellar masses below the peak, a feature of the model that
  should be verifiable by observations. The differential stellar mass
  function shows the second peak at $M_*\simeq 10^{6} \, M_\odot$
  first noted by \citet{Bose2018} and attributed to the abrupt change
  in the critical mass threshold caused by reionization. The second
  peak, however, is less prominent in our study because of the
  smoother transition in the critical threshold compared to that in
  \citet{Bose2018}.

\item Satellites settle onto orbits with apocentric radii that match,
  on average, the virial radius of the main halo at the time of first
  infall. The predicted satellite orbits are highly eccentric, with a
  median pericentre-to-apocentre ratio of $1$:$5$. The model also
  predicts a substantial fraction of luminous satellites with rather
  small pericentric radii. The MW has a dearth of satellites on such
  orbits, a disagreement that may be resolved once tidal stripping of
  stars is included in GALFORM, a calculation we defer to a future
  contribution.

\item Half of all Aq-A satellites (about $160$ objects) are expected
  to be contained within $\sim 50$ kpc from the main halo
  centre. Interestingly, only $\sim 48$ of those inhabit subhalos that
  are still resolved at $z=0$ in our highest resolution simulation;
  the rest are all orphans. There are only $22$ MW satellites inside
  that radius, meaning, that, in principle, there is no obvious
  difficulty accounting for the observed number of luminous satellites
  in the inner regions of the MW in our model.

\end{itemize}

It has recently been argued that the observed number of satellites in
the inner regions of the Milky Way is much higher than the number of
subhalos that remain self-bound in cosmological N-body simulations
\citep{Kelley2019,Graus2019}. These authors concluded that the MW data
can only be reconciled with $\Lambda$CDM if luminous galaxies populate
subhalos of extremely low mass, far below the H-cooling limit that
determines the critical mass threshold in our study.

We argue that this conclusion is incorrect, mainly as a consequence of
ignoring the important contributions of orphans to the simulated
satellite population.  Indeed, the opposite would seem true: our model
predicts far more orphan satellites in the inner regions compared to
the relatively small number of known satellites near the centre of the
Milky Way, a discrepancy that might be explained by the neglect of
tidal disruption of the stellar component of satellites in our model.

Our study underscores the critical need for a proper treatment of the
orphan satellite population in order to provide reliable predictions
for the full satellite population of the Milky Way down to the
ultrafaint regime. This is particularly important since current and
upcoming deep observational surveys (e.g. DESI, Vera Rubin Observatory
LSST, Roman, Euclid) will survey galaxies in our neighbourhood to unprecedented
depth. 

\section*{Acknowledgements}
We thank the referee for their constructive report which has helped improve this manuscript.
ISS and CSF acknowledge support from the European Research Council
(ERC) Advanced Investigator grant to C.S. Frenk, DMIDAS (GA 786910)
and from the Science and Technology Facilities Council [ST/P000541/1]
and [ST/X001075/1].
The simulations for the Aquarius Project were carried out at the
Leibniz Computing Centre, Garching, Germany, at the Computing Centre
of the Max-Planck-Society in Garching, at the Institute for
Computational Cosmology in Durham, and on the ‘STELLA’ supercomputer
of the LOFAR experiment at the University of Groningen.  This work
used the DiRAC@Durham facility managed by the Institute for
Computational Cosmology on behalf of the STFC DiRAC HPC Facility
(www.dirac.ac.uk). The equipment was funded by BEIS capital funding
via STFC capital grants ST/K00042X/1, ST/P002293/1, ST/R002371/1 and
ST/S002502/1, Durham University and STFC operations grant
ST/R000832/1. DiRAC is part of the National e-Infrastructure.  ISS
thanks M.Lovell for the background image data for Fig.~\ref{FigMap}.
JFN acknowledges the hospitality of MPA in Garching and DIPC in Donostia-San
Sebasti\'an during the completion of this work.

\section*{Data Availability}

The simulation data underlying this article may be shared upon
reasonable request to the Virgo Consortium's steering committee.



\bibliographystyle{mnras}
\bibliography{archive} 

\appendix

\section{MW satellite samples used for comparison}\label{sec:appMWdata}

We compare our results to available data for MW satellites,
specifically to a sample of $66$ satellite candidates located within
300 kpc from the MW centre:

*Cetus2,
\textbf{*Columba1}, *DESJ0225+0304, \textbf{*Draco2}, *Eridanus3,
\textbf{*Grus1}, \textbf{*Grus2}, \textbf{*Horologium1},
\textbf{*Horologium2}, *Indus1, *Indus2, \textbf{*Pegasus3},
\textbf{*Phoenix2}, *Pictor1, \textbf{*Reticulum2},
\textbf{*Reticulum3}, \textbf{*Sagittarius2}, \textbf{*Triangulum2},
\textbf{*Tucana3}, \textbf{*Tucana4}, \textbf{*Tucana5},
\textbf{Antlia2}, \textbf{Aquarius2}, \textbf{Bootes1},
\textbf{Bootes2}, \textbf{Bootes3}, Bootes4, \textbf{Bootes5},
\textbf{CanesVenatici1}, \textbf{CanesVenatici2}, \textbf{Carina},
\textbf{Carina2}, Carina3, Centaurus1, Cetus3, \textbf{ComaBerenices},
\textbf{Crater2}, Delve2, \textbf{Draco},
         Eridanus4,   \textbf{Fornax},   \textbf{Hercules},   \textbf{Hydra2},
         \textbf{Hydrus1},   Leo1,   \textbf{Leo2},   \textbf{Leo4},   \textbf{Leo5},   LeoMinor1,   
         \textbf{Pegasus4},   Phoenix,   Pictor2,   \textbf{Pisces2},   \textbf{Sculptor},
         \textbf{Segue1},   \textbf{Segue2},   \textbf{Sextans1},   \textbf{Tucana2},   \textbf{UrsaMajor1},
         \textbf{UrsaMajor2},   \textbf{UrsaMajor3/UNIONS1},   \textbf{UrsaMinor},   Virgo1,
         \textbf{Willman1},  LMC,  SMC.

         Names preceded by an asterisk indicate objects not
         spectroscopically confirmed as dwarf galaxies. Names in bold
         indicate $47$ objects with available kinematic data (i.e.,
         line-of-sight velocity and proper motions), for which we
         derive pericentre and apocentre distances at $z=0$.  These
         have been computed by integrating backwards the $z=0$
         3D position and velocity of the object within the
         \citet{Cautun2020} MW potential, extended beyond 300 kpc
         assuming an NFW profile with $M_{200}=10^{12}$ M$_\odot$.
         Three-dimensional 
         phase-space information has been computed from RA, Dec,
         $(m-M)$, line-of-sight velocity and proper motion data, as
         listed in the latest update of \citet{McConnachie2012}'s
         Nearby Dwarf Galaxy database (see Sec.~\ref{sec:MWdata}).  We
         only consider galaxy candidates with derived apocentres
         smaller than $600$ kpc. These are listed in
         Table~\ref{tab:MWperiapo} and plotted in Fig.~\ref{FigOrbits}.
         (Note we exclude the LMC and SMC from this subsample and from
         Fig.~\ref{FigOrbits}).

\begin{table}
	\centering
	\caption{Derived $z=0$ pericentre and apocentre distances for
          MW satellite candidates with available kinematic data.  For
          details see Sec.~\ref{sec:MWdata} and  Appendix~\ref{sec:appMWdata}. 
	Only galaxy candidates with derived apocentres smaller than
        $600$ kpc are listed below and plotted in
        Fig.~\ref{FigOrbits}. 
	}
	\label{tab:MWperiapo}
			\begin{tabular}{l l l} 
		\hline
		 Name  &  $r_{\rm peri}$(kpc) &   $r_{\rm apo}$(kpc)  \\
		\hline
		\hline
*Columba1    &    180.3    &    476.1 \\
*Draco2    &    20.7    &    140.7 \\
*Grus1    &    22.9    &    290.6 \\
*Grus2    &    21.0    &    72.4 \\
*Horologium1    &    71.8    &    105.9 \\
*Horologium2    &    37.5    &    79.9 \\
*Pegasus3    &    10.8    &    215.5 \\
*Phoenix2    &    78.3    &    319.6 \\
*Reticulum2    &    23.8    &    55.4 \\
*Reticulum3    &    19.1    &    113.6 \\
*Sagittarius2    &    54.4    &    194.2 \\
*Triangulum2    &    13.7    &    155.1 \\
*Tucana3    &    3.1    &    50.2 \\
*Tucana4    &    31.8    &    72.5 \\
*Tucana5    &    25.6    &    115.1 \\
Antlia2    &    71.7    &    149.8 \\
Aquarius2    &    85.6    &    125.1 \\
Bootes1    &    39.9    &    90.7 \\
Bootes2    &    38.7    &    307.7 \\
Bootes3    &    6.8    &    142.5 \\
Bootes5    &    20.3    &    113.8 \\
CanesVenatici1    &    60.1    &    270.3 \\
CanesVenatici2    &    10.9    &    196.6 \\
Carina    &    106.2    &    140.9 \\
Carina2    &    28.3    &    407.0 \\
ComaBerenices    &    44.4    &    139.2 \\
Crater2    &    44.5    &    149.1 \\
Draco    &    48.1    &    107.7 \\
Fornax    &    97.0    &    166.1 \\
Hercules    &    68.3    &    336.1 \\
Hydra2    &    35.7    &    195.2 \\
Hydrus1    &    25.3    &    247.5 \\
Leo2    &    125.2    &    242.2 \\
Leo4    &    26.8    &    154.8 \\
Leo5    &    44.3    &    209.1 \\
Pegasus4    &    31.3    &    92.2 \\
Pisces2    &    17.6    &    200.0 \\
Sculptor    &    65.1    &    129.2 \\
Segue1    &    20.4    &    60.8 \\
Segue2    &    18.2    &    45.3 \\
Sextans1    &    79.4    &    231.9 \\
Tucana2    &    35.3    &    257.1 \\
UrsaMajor1    &    51.2    &    101.6 \\
UrsaMajor2    &    35.9    &    91.6 \\
UrsaMajor3/UNIONS1    &    13.4    &    38.6 \\
UrsaMinor    &    50.6    &    99.4 \\
Willman1    &    20.4    &    43.3 \\
		\hline
	\end{tabular}
	
\end{table}

\section{Abundance of satellites as a function of the redshift of reionization}\label{sec:appzrei}
\begin{figure*}
\centering
\includegraphics[width=0.9\linewidth]{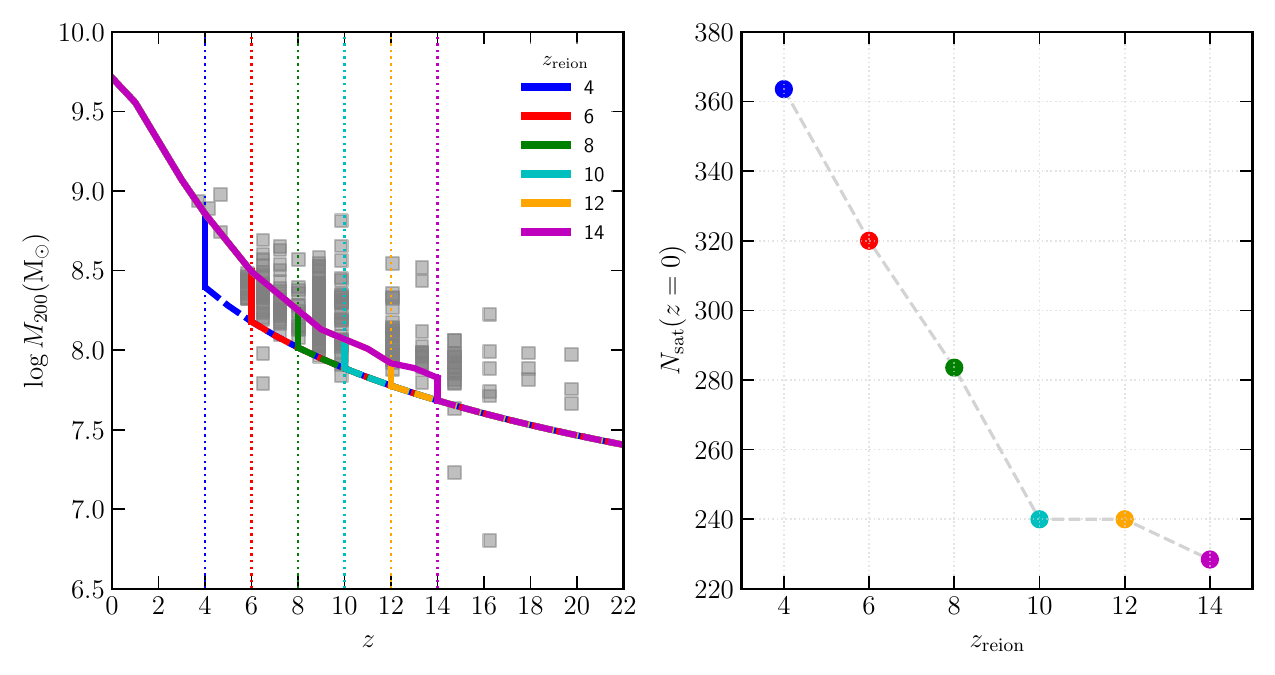}
\caption{ \textit{Left}: as the left panel of Fig.~\ref{FigGalform},
  but showing the redshift evolution of the critical mass when
  assuming different redshifts for reionization, as given in the
  legend.  We refer the reader to Sec.~\ref{SecGalform} for details of
  GALFORM modelling. For reference, grey datapoints show the results
  for our fiducial $z_{\rm reion}=6$ model (the same points as in
  Fig.~\ref{FigGalform}).  \textit{Right}: number of satellite
  galaxies at $z=0$ predicted by the model assuming different
  redshifts for reionization. The colour-coding is as in the left
  panel.  }
 \label{fig:appB1}
\end{figure*}

\begin{figure*}
\centering
\includegraphics[width=0.9\linewidth]{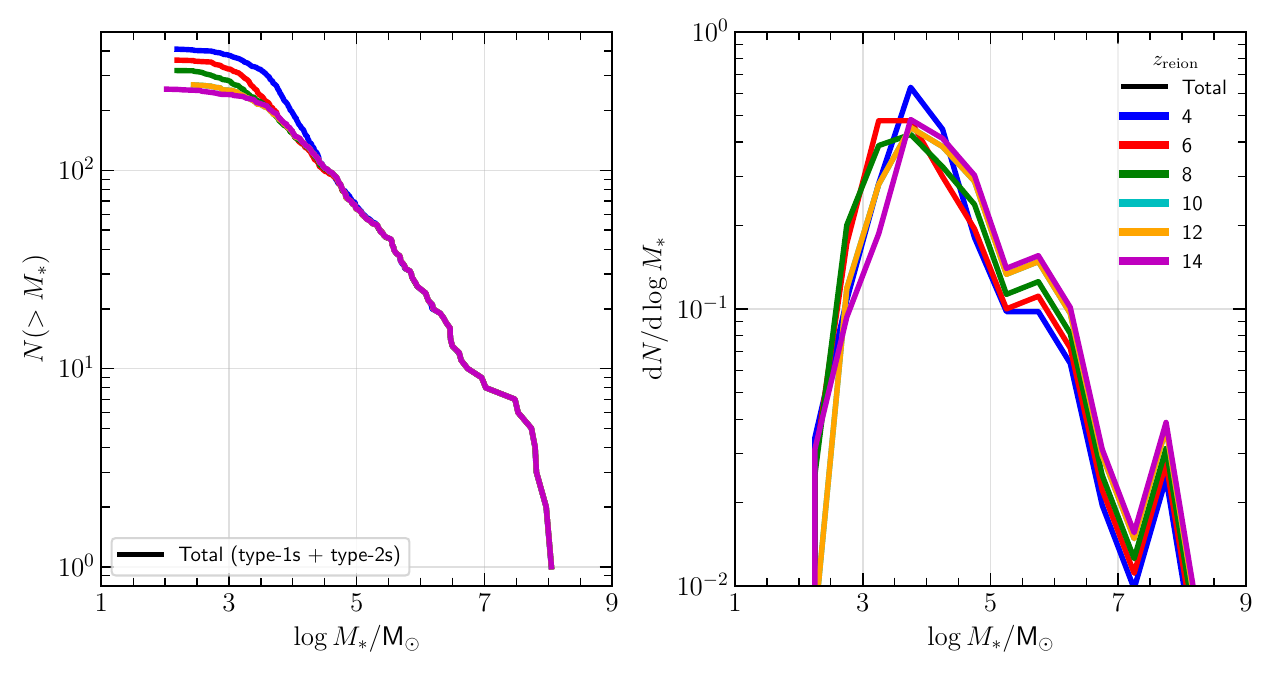}
\caption{ Cumulative (left) and differential (right) satellite
  luminosity function predicted by GALFORM with different assumed
  redshifts for reionization, as given in the legend.  }
 \label{fig:appB1b}
\end{figure*}

In this section we quantify the differences in the number of
satellites expected within $300$ kpc of the centre of the Aq-A-L1 host
as a function of the assumed redshift of reionization.  As explained
in Sec.~\ref{SecGalform}, gas cooling and subsequent star formation is
allowed in halos that exceed a critical mass, $M_{\rm crit}$, that
evolves with redshift, essentially following the virial mass of a halo
in which atomic hydrogen can cool.  This evolution shows a
discontinuity towards larger masses after reionization, when the gas
plasma is heated to $\sim 2\times10^4$ K.

The left panel of Fig.~\ref{fig:appB1} illustrates the redshift
evolution of $M_{\rm crit}$ for different values of $z_{\rm
  reion}$. Gray data points show results from our fiducial,
$z_{\rm reion}=6$, case, shown in Fig.~\ref{FigGalform}.  The earlier
$z_{\rm reion}$, the lower the halo mass at which the jump in
$M_{\rm crit}$ occurs, and the lower the number of expected luminous
satellites.  Indeed, the right panel of Fig.~\ref{fig:appB1} shows
that the number of satellites around Aq-A-L1 is heavily dependent on
$z_{\rm reion}$: while for $z_{\rm reion}=6$ we found $320$
satellites, for $z_{\rm reion}=10$ we find $230$, roughly $\sim30$\%
fewer.\footnote{These numbers have been corrected to account for the
  estimated percentage, $\sim11$\%, of `spurious' low-mass subhalos in which
  GALFORM may have mistakenly allowed stars to form due to
  `central-swapping' or other merger tree artifacts.}

This analysis shows that the number of satellites is not a robust
prediction of the model, so long as the redshift of reionization
remains uncertain.  On the other hand, while the total value of
$N_{\rm sat}$ may change when varying $z_{\rm reion}$, the associated
differences in the satellite luminosity function affect only the low
mass end, $M_{\rm star}<10^{3.5}$ M$_\odot$ (see
Fig.~\ref{fig:appB1b}).  Indeed, the shapes of both the cumulative and
differential luminosity functions remain essentially unaffected.  This
is a result of the luminosity function being determined by the
underlying subhalo $V_{\rm peak}$ function, which is a fundamental
prediction of $\Lambda$CDM, combined with the
$M_{\rm star}-V_{\rm peak}$ relation, which is a robust prediction of
the GALFORM model.


\bsp	
\label{lastpage}
\end{document}